\documentclass[pra,
a4paper,
showpacs,twocolumn,superscriptaddress]{revtex4}
\usepackage{amssymb}
\usepackage{amsmath}
\usepackage{amsfonts}
\usepackage{graphicx}
\usepackage{bm}
\usepackage{color}
\usepackage{multirow}
\usepackage{natbib}
\usepackage{hyperref}
\usepackage[normalem]{ulem}

\usepackage{verbatim}
\def \be {\begin{equation}}
\def \ee {\end{equation}}
\def \bea {\begin{align}}
\def \eea {\end{align}}
\def \p {\partial}
\def \BEA {\begin{eqnarray}}
\def \EEA {\end{eqnarray}}

\def \BC {\begin{cases}}
\def \EC {\end{cases}}

\def \be {\begin{equation}}
\def \ee {\end{equation}}
\def \bea {\begin{align}}
\def \eea {\end{align}}
\def \p {\partial}
\def \BEA {\begin{eqnarray}}
\def \EEA {\end{eqnarray}}

\def \BC {\begin{cases}}
\def \EC {\end{cases}}

\begin{document}

\title{Quantum elasticity of graphene: Thermal expansion coefficient and specific heat}

\author{I. S.~Burmistrov}
\affiliation{L. D.~Landau Institute for Theoretical Physics, Kosygina
  street 2, 119334 Moscow, Russia}
\affiliation{Laboratory for Condensed Matter Physics , National Research University Higher School of Economics, 101000 Moscow, Russia}
\author{I. V.~Gornyi}
\affiliation{L. D.~Landau Institute for Theoretical Physics, Kosygina
  street 2, 119334 Moscow, Russia}
\affiliation{Institut f\"ur Nanotechnologie,  Karlsruhe Institute of Technology,
76021 Karlsruhe, Germany}
\affiliation{\mbox{Institut f\"ur Theorie der kondensierten Materie,  Karlsruhe Institute of
Technology, 76128 Karlsruhe, Germany}}
\affiliation{A. F.~Ioffe Physico-Technical Institute,
194021 St.~Petersburg, Russia}
\author{ V. Yu.~Kachorovskii}
\affiliation{L. D.~Landau Institute for Theoretical Physics, Kosygina
  street 2, 119334 Moscow, Russia}
\affiliation{Institut f\"ur Nanotechnologie,  Karlsruhe Institute of Technology,
76021 Karlsruhe, Germany}
\affiliation{\mbox{Institut f\"ur Theorie der kondensierten Materie,  Karlsruhe Institute of
Technology, 76128 Karlsruhe, Germany}}
\affiliation{A. F.~Ioffe Physico-Technical Institute,
194021 St.~Petersburg, Russia}

\author{M. I.~Katsnelson}
\affiliation{Radboud  University,  Institute  for  Molecules  and  Materials,  NL-6525AJ  Nijmegen, The  Netherlands}
\author{A. D.~Mirlin}
\affiliation{L. D.~Landau Institute for Theoretical Physics, Kosygina
  street 2, 119334 Moscow, Russia}
\affiliation{Institut f\"ur Nanotechnologie,  Karlsruhe Institute of Technology,
76021 Karlsruhe, Germany}
\affiliation{\mbox{Institut f\"ur Theorie der kondensierten Materie,  Karlsruhe Institute of
Technology, 76128 Karlsruhe, Germany}}
\affiliation{Petersburg Nuclear Physics Institute, 188300, St.Petersburg, Russia}

\date{\today}

\begin{abstract}
We explore thermodynamics of a quantum membrane, with a particular application to suspended graphene membrane and with a particular focus on the thermal expansion coefficient.
We show that an interplay between quantum and classical
anharmonicity-controlled fluctuations  leads to unusual elastic properties of the membrane.   The  effect of quantum  fluctuations   is  governed by the  dimensionless coupling constant, $g_0 \ll 1$, which vanishes in the classical limit ($\hbar \to 0$) and is equal to $\simeq 0.05$ for graphene.  We demonstrate that  the thermal expansion coefficient $\alpha_T$ of  the membrane  is negative and
remains nearly constant down to  extremely  low temperatures, $T_0\propto  \exp (-2/g_0)$.
We also find that $\alpha_T$ diverges in the classical limit:  $\alpha_T \propto - \ln(1/g_0)$ for $g_0 \to 0$.
For graphene parameters, we estimate the value of the thermal expansion coefficient as  $\alpha_T \simeq  - 0.23\:{\rm eV}^{-1}$,   which applies below the temperature $T_{\rm uv} \sim g_0 \varkappa_0 \sim 500$\:K  (where $\varkappa_0 \sim 1$\:eV is the bending rigidity) down to  $T_0 \sim 10^{-14}$\:K.
For $T<T_0$, the thermal expansion coefficient slowly (logarithmically) approaches zero with decreasing temperature.
This behavior is
surprising since typically the thermal expansion coefficient goes to zero as a power-law function.
We discuss possible experimental consequences of this anomaly.
We also evaluate classical and quantum contributions to  the specific heat of the membrane and investigate the behavior of the Gr\"uneisen parameter.
    \end{abstract}
\maketitle

\section{Introduction\label{sec.I}}

 The thermal expansion coefficient $\alpha_T$ is one of the most important thermodynamic  characteristics of  any material. It is well known  that    $\alpha_T$ is determined by an anharmonicity  of interatomic potentials binding atoms into a crystalline lattice \cite{peierls,cowley1,cowley2,leib,KT2002,encyc}. Indeed,  for a harmonic oscillator, the averaged displacement of the coordinate  equals zero independently of the  oscillation amplitude.  By contrast,  for an anharmonic oscillator, an averaged displacement  depends on the oscillation amplitude and, consequently, on temperature.

 For the most of materials, the thermal expansion coefficients are positive, $\alpha_T>0$.
 On the other hand,  some exotic systems---including, in particular, complex metal oxides, polymers, and zeolites---are known to contract upon heating [\onlinecite{NEC}]. The  interest to   materials with $\alpha_T<0$  is motivated, in particular,   by  the desire   to fabricate a composite structure fully compatible with conventional semiconductor nanotechnology  and  having zero thermal expansion coefficient.  A certain  progress in this direction is connected with recent observation that carbon nanotubes [\onlinecite{cn66,cn67,cn68,cn69}] might demonstrate  negative thermal expansion (NTE). However, the measured  effect  was relatively small and observed in a not too wide  temperature interval.  It is also worth  noting that  materials  demonstrating  NTE at very low temperatures (below 0.1 K) are unknown so far.

The goal of this paper is to demonstrate that  graphene, a famous two-dimensional (2D) material that  has been attracting  enormous interest in last decade
[\onlinecite{Geim,Geim1,Kim,geim07,graphene-review,review-DasSarma,review-Kotov,book-Katsnelson,book-Wolf,book-Roche}],
 shows NTE with an approximately constant $\alpha_T$  for all experimentally accessible temperatures: from  very  high (a few hundreds of Kelvin)  temperatures   down  to  extremely  low temperatures, $T_0 \sim 10^{-14}$ K.   Only at exponentially small temperatures,   $T < T_0$, the thermal expansion coefficient $\alpha_T$  goes to zero.  Since  measurements of the elasticity of free-standing graphene have recently become accessible to experimental techniques [\onlinecite{blees15,lopez-polin15,nicholl15}], the prediction of an approximately temperature-independent negative $\alpha_T$ can be verified experimentally. Our consideration is applicable also to other two-dimensional materials.

The fact that NTE is a natural property of layered  and 2D materials due to existence of bending (flexural acoustic) mode which is anomalously sensitive to the deformation has been recognized long time ago \cite{lifshitz} and dicussed in a number of   recent publications \cite{book-Katsnelson,marzari,Katsnelson1,ACR}). In particular, graphite is known to have NTE at not too high temperatures \cite{Steward}, which may be explained quantitatively within a quasiharmonic theory \cite{cowley1,cowley2,leib,KT2002,encyc} via first-principles calculations of phonon spectra and Gr\"uneisen parameters \cite{marzari}. The same calculations predicted that graphene should have NTE at any temperatures. More recently, atomistic simulations of graphene \cite{Zakh} confirmed the NTE at moderately high temperatures and also showed that the thermal expansion coefficient changes sign with growing temperature due to anharmomic effects beyond the quasiharmonic approximation. Qualitatively similar results were obtained in Refs.~\cite{peet1,peet2}.
Experiments confirm that the thermal expansion coefficient of graphene at room temperature is negative, with the absolute value as large as 0.1 ${\rm eV}^{-1}$ \cite{Bao}.  With the temperature increase up to 400 $K$,  $\alpha_T$ decreases in absolute value \cite{Yoon}, which may be considered as a partial confirmation of the prediction \cite{Zakh}. At very high temperatures, the thermal expansion coefficient of graphene is definitely positive but its precise measurement is difficult  \cite{Boer}.

The main focus of the present paper is on the range of relatively low temperatures.  The behavior of the thermal expansion coefficient of graphene (or, more generally, of a 2D membrane) in this regime represents a challenging theoretical problem.
The microscopic Gr\"uneisen parameter for the bending mode is divergent at the phonon wave vector $q \rightarrow 0$ in quasiharmonic approximation as $-1/q^2$ (see Eq.~4 of \cite{Katsnelson1}), which means that relevant phonons determining the thermal expansion are always classical (i.e, their energy is smaller than the temperature). As a result, within the quasiharmonic approximation,  the thermal expansion  coefficient remains constant down to  arbitrary low temperatures \cite{Katsnelson1}.
This is in stunning contrast with the conventional  behavior, $\alpha_T \rightarrow  0$   at $T \rightarrow  0,$ which is usually associated  with the third law of thermodynamics in view of the identity \cite{landau}
 \begin{equation}
 \left(\frac{\partial V}{\partial T}\right)_P = - \left(\frac{\partial S}{\partial P}\right)_V
 \label{third}
 \end{equation}
where $V,P,T,S$ are volume, pressure, temperature, and entropy, respectively.

Clearly, one may expect that quantum effects modify the low-$T$ behavior of $\alpha_T.$
Quantum corrections to thermodynamic properties have been calculated in Ref.\onlinecite{Katsnelson2}. Based on perturbative analysis, it was suggested that classical theory becomes inapplicable already at reasonably high temperatures, about 70--90 $K$ and that the thermal expansion coefficient goes to zero
 in a conventional way---i.e., as a power law---at $T \rightarrow  0.$
 Here we will show that the situation is, in fact, much more exotic and the classical expression for the thermal expansion coefficient \cite{Katsnelson1} remains valid, with some logarithmic corrections, till very low temperatures.  The thermal expansion coefficient does go to zero at zero temperature
 but slower than any power of the temperature.

 In this paper, we explore systematically thermodynamic properties of a graphene membrane, with a particular focus on the thermal expansion coefficient $\alpha_T$, in the whole range of temperatures. We show, in particular, that the 
 behavior of $\alpha_T$  in the limit $T\rightarrow 0$ is connected with quantum effects characterized by a dimensionless coupling constant $g$  [definition of  $g$ is given by  Eq.~\eqref{g0} below], which vanishes in the classical limit $\hbar \to 0 .$

As a one of the most important results of the paper,   we demonstrate that the thermal expansion coefficient of the membrane
  remains  negative  and nearly constant  down  to all realistic temperatures.
   For graphene parameters this constant  value is about
  $ - 0.23\:{\rm eV}^{-1}.$
   For a generic membrane, we find that  in the limit  of weak coupling,  the  value of $\alpha_T$ depends logarithmically on the bare value of the coupling  $g_0$ (the value of $g$ at the  atomic scales):
 $$\alpha_T \propto -\ln(1/g_0), \,\,\, \text{for}\,\,\,g_0 \ll 1, $$
and consequently diverges in the classical limit $\hbar \to 0 .$
         We also demonstrate that, with decreasing temperature, $\alpha_T$ starts to approach zero only when $T$ drops below an exponentially small temperature scale $T_0$
\be
T_0 \sim g_0 \varkappa_0 e^{-2/g_0}.
\label{Tuv}
\ee
Here $\varkappa_0$ is the bare value of the bending rigidity of the membrane (equal to $\simeq 1$eV for graphene). Furthermore, the decay of $|\alpha_T|$ at such exponentially low temperatures is logarithmically slow.

We also calculate the  specific heat.  In particular, we show that the leading contribution to both $C_V$ and $C_P$ is determined by classical effects, while the difference   $C_P -C_V$  is proportional to $g_0$ and, thus, has a purely quantum nature.

The outline of the paper is as follows. Section \ref{secIa} contains a qualitative discussion of the physics of fluctuation-induced elasticity of a membrane. In Sec. \ref{sec.II} we present the classical and quantum renormalization-group (RG) formalism as well as basic equations for the observables of interest.  In Sec. \ref{sec.III} we use the results of the preceding section to calculate the temperature dependence of the thermal expansion coefficient and of the specific heat. Our results are summarized in Sec. \ref{con}. Various technical aspects of our analysis are presented in   Appendices \ref{F-S}--\ref{app-D}.

\section{Fluctuation-induced elasticity of a membrane}
\label{secIa}

A free-standing graphene is a remarkable example of a 2D crystalline membrane \cite{book-Katsnelson,ACR}. Elastic properties of  such a membrane are characterized by the bending rigidity $\varkappa$ (which is quite high for graphene,   ${\varkappa\simeq 1}$~eV),  the Young  modulus $Y$  and the bulk modulus $B$:
\be
Y=\frac{4\mu (\mu+\lambda)}{2\mu+\lambda},\quad B=\mu+\lambda,
\ee
where $\mu$ and $\lambda$ are the Lame coefficients (for graphene  $\lambda \simeq$  2 eV$\cdot$\AA$^{-2}$ and $\mu \simeq$ 10 eV$\cdot$\AA$^{-2}$, see Ref.\onlinecite{Zakh}).
A distinct feature of a free-standing 2D membrane  is the
existence of out-of-plane phonon modes---flexural phonons (FP) [\onlinecite{lifshitz,book-Katsnelson,ACR,Nelson}].
In contrast to in-plane phonons with the conventional linear dispersion,
the FP are very soft, $\omega_{ q} \propto q^2$. As a consequence,
the out-of-plane thermal fluctuations are unusually strong and tend to destroy a membrane by
driving it into the crumpled phase [\onlinecite{Nelson}].

\begin{figure}[t]
\centerline{\includegraphics[width=0.3\textwidth]{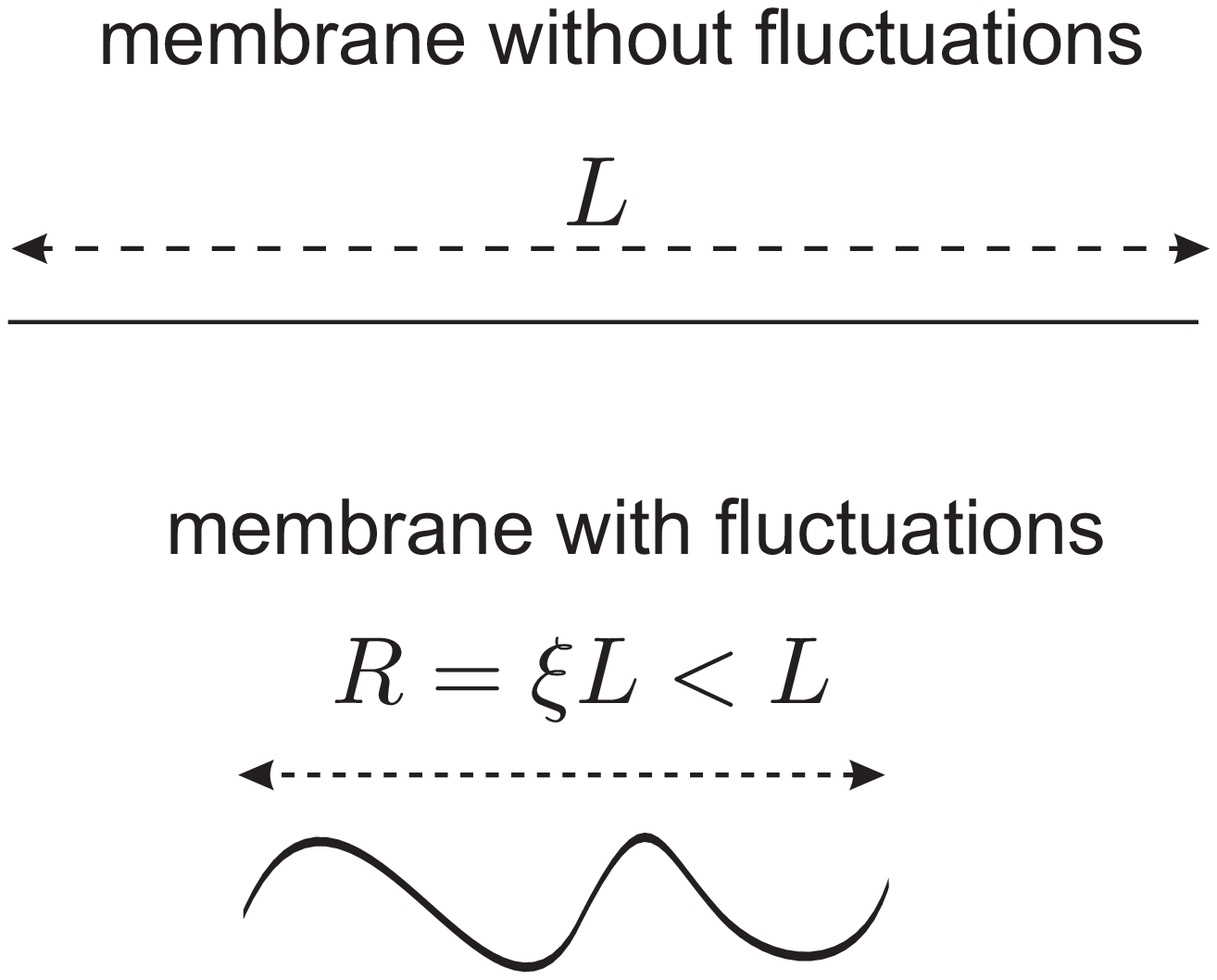} }
\caption{Temperature-induced shrinking of a graphene membrane.
Due to classical and quantum fluctuations, the  membrane shrinks  in  the  longitudinal direction (bottom), as  compared to the membrane without fluctuations (top).
    }
\label{Fig1}
\end{figure}

The tendency of the membrane to crumpling is at the heart of the mechanism leading to NTE. Let us explain this in more detail. The vector $\mathbf r $ describing point at the membrane surface depends on  the $2D$ coordinate $\mathbf x$ that parametrizes the
membrane and
 can be split into three terms
\be
\mathbf r=\xi \bm{x}+ \bm{u} +\bm{h},
\label{r}
\ee
where
vectors
$\bm{u}=\bm{u}(\bm{x},t),~ \bm{h} =\bm{h}(\bm{x},t) $
represent in-plane and out-of-plane phonon fields, respectively.
The  global stretching  factor $\xi$  is
equal to unity  at zero temperature  but
gets reduced  with increasing $T$ due to  out-of-plane fluctuations  (see  Fig.~\ref{Fig1}),  becoming zero at the crumpling transition temperature $T_{\rm cr}.$   Remarkably,  the thermal expansion of membrane  is nonzero even if one neglects  the  anharmonicity of in-plane and out-of-plane modes as well as anharmonic   coupling between them [\onlinecite{lifshitz}].    Within such an   approximation  the only  relevant anharmonicity  is due to the coupling of global stretching and FP.  Treating FP as classical random fields   yields [\onlinecite{my-crump}]
\begin{equation}
\xi^2=1-  \langle \p_\alpha \bm{h} \p_\alpha \bm{h} \rangle/2,
\end{equation}
which results in a negative, logarithmically divergent value of $\alpha_T =   L^{-2} {dA}/{dT}$:
\be
\alpha_T=\frac{\p \xi^2}{ \p T} = - \frac{1}{4\pi \varkappa} \ln\left(\frac{L}{L_{\rm uv}}\right) .
\label{alpha-harmonic}
\ee
Here, $A=\xi^2L^2 $  is the sample area,   $L$ is the system size at  zero temperature, and $L_{\rm uv}$   is the ultraviolet cutoff  of the order of the lattice constant.
Evidently, if  this equation were fully correct,  the membrane would not  exist  in the thermodynamic limit,  $L\to \infty$.  Actually,  the anharmonic coupling between $\bm{u}$ and $\bm{h}$ fields  leads to a renormalization of the bending rigidity, which becomes momentum-dependent, $\varkappa  \rightarrow \varkappa_q$ and flows to infinity, $\varkappa_q   \to \infty$,  for $q\to 0$ [\onlinecite{Nelson,Nelson0,Crump1,NelsonCrumpling,david1,buck,Aronovitz89,david2,lower-cr-D2,d-large,disorders,disorder-imp,Gompper91,
RLD,Doussal,disorders-Morse-Grest,RLD1,Bowick96,my-crump,book-Katsnelson,ACR}].
This cures the  logarithmic divergency of $\alpha_T$ with the system size, thus yielding a finite value for the crumpling transition temperature, $T_{\rm cr} < \infty$, for $L\to \infty$. Below $T_{\rm cr}$ one gets [\onlinecite{my-crump,my-hooke}]
\begin{equation}
\xi^2 = 1- T/T_{\rm cr},
\end{equation}
and, consequently, a finite negative value of  the  thermal expansion coefficient,
$\alpha_T= - 1/T_{\rm cr}.$

These arguments indicate that the role of anharmonicity  in  a free-standing 2D membrane is remarkably different as compared to the case of a 3D crystal (or to the case of  graphene on substrate). Indeed, in the 3D case, the anharmonic coupling between phonons determines a nonzero value of $\alpha_T$ (which is zero in the harmonic approximation).  In contrast, in a free-standing 2D membrane, such coupling  leads to a suppression of an infinite value of $\alpha_T$ predicted within the harmonic description of in-plane and flexural modes down to a finite value.

In order to find  $T_{\rm cr}$ and $\alpha_T$, one should investigate the renormalization from the ultraviolet energy scale down to the infrared scale. Such renormalization was  intensively discussed more than two decades ago
[\onlinecite{Nelson,Nelson0,Crump1,NelsonCrumpling,david1,buck,Aronovitz89,david2,lower-cr-D2,d-large,disorders,disorder-imp,Gompper91,
RLD,Doussal,disorders-Morse-Grest,RLD1,Bowick96}]  in the purely classical approximation (under assumption that the temperature is higher than the relevant frequencies of in-plane and flexural modes)
in connection with  biological membranes, polymerized layers, and inorganic surfaces.
The interest to this topic has been renewed
more recently~[\onlinecite{eta1,Gazit1,Hasselmann,kats1,kats2,Katsnelson2,kats-erratum,kats3,scaling,kats4,Katsnelson3,moh}]
after discovery of graphene.
It was found [\onlinecite{Nelson0,Crump1,NelsonCrumpling,david1,Aronovitz89,buck,david2,lower-cr-D2,Doussal}] that the anharmonic coupling of
in-plane and out-of-plane phonons  leads  to
 power-law renormalization of the bending rigidity, $\varkappa\to\varkappa_q$, where
\be
\varkappa_q \simeq \varkappa  \left(\frac{q_*}  {q}\right)^{\eta},~~\text{for}~~q \ll q_*,
\label{kappa-q}
\ee
with a certain critical exponent $\eta$. Here the momentum scale (see e.g. Eq.~(154) of Ref. [\onlinecite{my-crump}])
\be
q_*  = \sqrt{ \frac{3 d_{\rm c} Y T }{32\pi \varkappa^2}} \sim \frac{\sqrt{d_{\rm c }\mu T}}{ \varkappa}
\label{q-star}
\ee
is the inverse Ginzburg length which separates the regions of conventional ($q>q_*$) and fractal ($q<q_*$) scaling \cite{Nelson,Nelson0,book-Katsnelson,ACR}.
In other words, the anharmonicity of flexural modes becomes important at $q<q_*$. The last expression in Eq.~(\ref{q-star}) is an order-of-magnitude estimate where we have discarded numerical coefficients as well as a difference in values of the elastic moduli,   $Y\sim B\sim \mu\sim \lambda$. We will present such estimates also in several cases below in order to emphasize the scaling of observables with parameters of the problem.

The critical exponent $\eta$ was determined within several
approximate analytical schemes [\onlinecite{david1,david2,Aronovitz89,Doussal,eta1}]. In particular, for a 2D  membrane embedded into a space of large dimensionality $d \gg 1$, one can find  analytically  $\eta=2/d_{\rm c} \ll1,$ where $$d_{\rm c}=d-2.$$
Numerical simulations for physical  2D membrane embedded in 3D space ($d_{\rm c}=1$) yield $\eta=0.60 \pm 0.10$ [\onlinecite{Gompper91}]
and $\eta=0.72 \pm 0.04$ [\onlinecite{Bowick96}]. Atomistic Monte Carlo simulations for graphene gives the value $\eta \approx 0.85$ \cite{ACR,scaling}; approximately the same value has been derived via functional renormalization group approach \cite{eta1}.

In the limit $d_{\rm c}  \gg1,$ the critical temperature of the crumpling transition can be  also calculated analytically  within a classical approximation, yielding  [\onlinecite{my-crump,my-hooke}]
\be
T_{\rm cr}=\frac{4\pi \eta \varkappa}{d_{\rm c}} = \frac{8\pi \varkappa}{d_{\rm c}^2}.
\label{Tcr0}
\ee
As a consequence, the
thermal expansion coefficient  is negative and independent of temperature
\be
\alpha_T^ {d_{\rm c} \to \infty } = -\frac{d_{\rm c}}{4\pi \eta \varkappa}.
\label{alphaT-d-infty}
\ee
Evidently, one expects that   this result should fail at low enough temperatures due to the quantum effects.

Some aspects of this problem were discussed recently \cite{Katsnelson1,Katsnelson2,kats3,Katsnelson3}.
One scenario of the suppression of $\alpha_T$ is the emergence of a ``mass'' (term quadratic in momentum $q$) in the propagator of flexural phonons or, equivalently,  $1/q^2$ divergence of the effective bending rigidity  \cite{Katsnelson2}. This mass arises naturally within perturbative calculations \cite{Katsnelson2}.
However, as it will be shown below by an explicit calculation of the free energy,
it is nothing else as the full tension (as was pointed out in Ref. \onlinecite{kats3}) and thus is zero for a free membrane (without external stress).


As was shown in Ref.~\onlinecite{kats2}, quantum fluctuations  may also lead  to renozmalization of elastic coefficients.  This renormalization can be described \cite{kats2,kats-erratum}  in terms of a flow of the dimensionless  quantum    coupling constant
\be
g = \frac{3 (6+d_{\rm c})}{128\pi} \frac{ \hbar Y}{\rho^{1/2} \varkappa^{3/2}} \sim \frac{ \hbar \mu }{\rho^{1/2} \varkappa^{3/2}} ,
\label{g0}
\ee
where $\rho$ is the mass density of  membrane.  Since $g \propto \hbar,$ it vanishes in the classical limit $\hbar \to 0 .$ For graphene, the bare value $g_0$ of this constant   at the  ultraviolet scale, $q\sim q_{\rm uv}$, on the order of the lattice constant   is quite small,  $g_0 \simeq 1/20$.  Physically, this happens because $g_0$ contains $\rho$ and, consequently, the  atomic mass in the denominator.

In the sequel, we develop a theory of thermal expansion of an elastic membrane  that includes both classical and quantum effects. We show that  there is an additional contribution   to  $\alpha_T,$ which originates from
the region of momenta  $ q_*<q < q_*/\sqrt{ g}$ and is not taken into account in Eq.~\eqref{alphaT-d-infty}. This contribution is  logarithmically  large [$\propto -\ln(1/g)$] for small coupling $g.$
Quantum fluctuations originate from   $q>q_*/\sqrt g.$   Evidently, such interval of $q$ exists only when $q_*/\sqrt g <q_{\rm uv}.$ The temperature found from the condition $ q_*/\sqrt g  \sim q_{\rm uv}$ is given by $T_{\rm uv}\sim g_0\varkappa_0 $ ($\sim 500$ K for graphene). This temperature is determined by the bare value $g_0$ of the quantum coupling constant  and  plays the role of the Debye temperature.   For $T<T_{\rm uv}$ the problem becomes quantum in terms of statistics, i.e., some phonons are frozen out.     On the other hand,
the effect of quantum fluctuations (i.e., those with momenta in  the range  $ q_*/\sqrt g <q<q_{\rm uv}$) remains negligibly small
 in a  wide temperature interval $ T_0<T<T_{\rm uv}$, where $T_0 \sim  T_{\rm uv} \exp(-2/g_0) $.
 Therefore, from the point of view of fluctuation-induced renormalization, the problem remains classical down to the temperature $T_{\rm uv}$, which is exponentially small for $g_0\ll 1$.  Only at very low temperatures, $T<T_0$,   quantum fluctuations come into play and, as a result,  the  thermal expansion coefficient  gets logarithmically  suppressed.

Our goal in this paper is to develop a theory of thermal expansion valid for all temperatures in the range $T<T_{\rm uv}$
 where  the main contribution  to $\alpha_T$ originates from  the flexural phonons and therefore is negative. This requires a development of formalism incorporating both classical and quantum renormalization effects, which is the subject of the next Section.

\section{Formalism\label{sec.II}}

\subsection{Thermodynamics of an elastic membrane}
\label{sec2a}

We  consider a  generic $2D$ membrane embedded in the $d$-dimensional space ($d>2$).   The starting point of our analysis is the Lagrangian density
\BEA
\label{L}
&&{\cal L}(\left\{\mathbf r\right\})\!=\rho \dot{\mathbf r}^2   +\!\frac{\varkappa_0 }{2}(\Delta{\mathbf r})^2\!
\\
&&
+ \frac{\mu_0}{4}(\partial_\alpha{\mathbf r}\partial_\beta{\mathbf r}\!-\!\delta_{\alpha\beta})^2
\!+\!\frac{\lambda_0}{8}(\partial_\gamma{\mathbf r}\partial_\gamma{\mathbf r}\!-\!2)^2\!,
\nonumber
\EEA
which can be obtained from the general  gradient expansion of  elastic
energy  \cite{NelsonCrumpling}
by using a certain rescaling of coordinates (see discussion in Ref.~\onlinecite{my-crump}). The subscript $0$ in notations for  elastic coefficients in Eq.~(\ref{L}) means that these are bare  values at $q\simeq q_{\rm uv},$ where   $q_{\rm uv}$ is the ultraviolet cut off.
The $d-$dimensional vector $\mathbf r= \mathbf r(\mathbf x,\tau)$   is given by Eq.~\eqref{r}  with $\mathbf u=(u_1,u_2),~ \mathbf h =(h_1,...,h_{d_c}) $, while  $\dot{\mathbf r}= d \mathbf r /d \tau,$ where $\tau$ is imaginary  time.

The  strategy of calculations is as follows.   We assume that the ahrahmonic  phonon interaction leads to a renormalization of elastic coefficients  $\varkappa_0 \to \varkappa_q$, $\mu_0 \to \mu_q$,  $\lambda_0 \to \lambda_q$.   Hence, in order to calculate  the free energy, we  replace   Eq.~\eqref{L} with  a harmonic Lagrangian density containing renormalized  elastic moduli.    The details of calculation are relegated to  Appendice \ref{F-S} and \ref{anh}.    The obtained free energy has the form [see Eq.~(\ref{Fsigma0-app})]
\BEA
\frac{F}{L^2}&=& -\frac{\sigma^2}{2B_0} +\sigma (\xi^2-1)
 \label{Fsigma0}
\\ \nonumber
&+&
   \frac {d_{\rm c}}{2} \sum \limits _{\mathbf q\omega}
   \ln \left( \varkappa_q q^4 +\sigma q^2 + \rho \omega^2\right),
\EEA
where
 \be
  \label{sigma-def-1}
  \sigma = \frac{1}{L^2} \frac{\partial F}{\partial \xi^2},
  \ee
is  the external stress,  $\sum \limits _{\mathbf q\omega} $ stands for $T\sum\limits_\omega \int d^2\mathbf q/(2\pi)^2, $
  and  the summation $\sum \limits _\omega$  runs  over bosonic Matsubara frequencies. This approximation is in spirit of self-consistent phonon theory \cite{SCAILD} where, in analogy to Landau Fermi liquid theory for fermions, it is supposed that the entropy is renormalized by phonon-phonon interaction only via the change of their dispersion relation and phonon damping is neglected. One can assume that, within some numerical factors, this gives correct temperature dependences of all thermodynamic quantities.

 Since FP are  much softer than in-plane modes, we neglected contribution of in-plane modes in Eq.~\eqref{Fsigma0}. Although  this approximation  looks quite natural,  it needs some justification.   Indeed, as was found in Ref.~\onlinecite{Katsnelson2}, the  anharmonicity induces a  small (in adiabatic parameter) ultraviolet-divergent  contribution $\sigma_{1}$ (``built-in tension'') to the coefficient in front of the $q^2$ term in the propagator of FP.    Such  term    would       suppress $\alpha_T$ (in a power-law way) at low temperatures.    In fact, this term is exactly cancelled by another contribution, which   was overlooked in Ref.~\onlinecite{Katsnelson2}.   Technically,  this additional contribution arises due to the  coupling between in-plane phonons  and global stretching.    Here, we discuss the problem  on the qualitative level, relegating  the details of calculations  to Appendices \ref{F-S} and \ref{anh}.
     Substituting Eq.~\eqref{r} into strain tensor of the membrane
     $$u_{\alpha\beta}= \frac12 \left( \partial_\alpha{\mathbf r}\partial_\beta{\mathbf r}\!-\!\delta_{\alpha\beta}\right),$$
     and leaving only linear with respect to fluctuation terms, we find that, in the harmonic approximation, the strain tensor is proportional to the global stretching and to the gradient of in-plane deformations:
     \be
     u_{\alpha\beta}^{\rm harmonic}=
     \xi(\p_\alpha u_\beta+\p_\beta u_\alpha) /2.
\label{strain-harmonic}
 \ee
 The  energy of the in-plane fluctuations  $E_{\rm in-plane}$ is quadratic with respect to $u_{\alpha\beta}$ \cite{Nelson} and, as follows from Eq.~\eqref{strain-harmonic}, is proportional to $\xi^2.$  Hence, there exists an  contribution to the stress
 \be
 \delta \sigma=   \frac{ \langle E_{\rm in-plane} \rangle}{\xi^2 L^2},
 \label{delta-sigma }
 \ee
 where averaging is taken with the action corresponding to the Lagrangian \eqref{L}.  As shown in Appendixes \ref{F-S} and   \ref{anh}
 \be
 \delta \sigma=\sigma_1.
 \label{cancel}
 \ee
  This implies that  the quadratic-in-$q$ part of the self-energy of FP is given by
  $(\sigma +\sigma_1-\delta \sigma) q^2=\sigma q^2,$ where $\sigma$ is the external tension.  In other words, coupling of in-plane modes to the global stretching $\xi$  leads to additional contribution to the FP's self-energy   which  exactly cancels the  quadratic correction arising due to anharmonic coupling of in-plane modes with FP.

  This cancelation has a deep physical meaning. There are two different definitions of the tension $\sigma$. First, it can be obtained from the standard thermodynamic relation, as a derivative of the free energy with respect to system volume, see Eq.~\eqref{sigma-def-1}.
Second, the tension can be found as a
   coefficient in front of the quadratic-in-$q$ term in the FP propagator,
    \be
  \label{sigma-def-2}
  \sigma =  \left. \frac{\partial G_q^{-1}}{\partial (q^2)} \right |_{q^2 = 0}.
  \ee
 The equivalence of two definitions, while quite transparent physically, represents a very non-trivial  Ward identity  \cite{buck,lower-cr-D2}. We explicitly verify this identity within the one-loop RG analysis in Appendixes~\ref{F-S} and \ref{app-D}.

  In-plane modes  lead also to a small
    ultraviolet renormalization of $\xi$ which we neglect here (see Appendix \ref{anh} for more detail).

The relation between $\xi$ and $\sigma$ is found from the condition  $\p F/ \p \sigma =0,$ which yields, after summation over Matsubara frequencies,
\be
\xi^2-\xi_0^2= \frac{\sigma}{B_0 } + s(\sigma),
\label{xi}
\ee
where
\be
 \xi_0^2=1-\frac{d_{\rm c}}{8\pi\rho}\int \limits _0^{q_{\rm uv}} \frac{dq~ q^3}{\omega_q^0}\coth \left(\frac{\omega_q^0}{2T}\right) , \label{xi0}
\ee
is the longitudinal deformation of membrane in the absence of stress, while
\be\!s\!=\! \frac{d_{\rm c}}{8\pi\rho}\!\int \limits _0^{q_{\rm uv}} \!\! dq q^3\!\!\left[\!\frac{\coth \left({\omega_q^0}/{2T}\right)} {\omega_q^0} \!-\! \frac{\coth \left({\omega_q}/{2T}\right)} {\omega_q}\!\right]\!, \label{S}
 \ee
 is the stress-induced  correction ($s_{\sigma \to 0}=0$),  which leads, in particular, to anomalous Hooke's law \cite{my-hooke}. Here  $\omega_q$ and $\omega_q^0$ are FP frequencies for stressed and unstressed membrane, respectively,
 \be
 \omega_q=\sqrt{\frac{\varkappa_q q^4+\sigma q^2}{\rho}},\qquad \omega_q^0=\sqrt{\frac{\varkappa_q}{\rho}} ~q^2.
 \label{omega-q}
\ee

\begin{figure}[t]
\centerline{\includegraphics[width=0.5\textwidth]{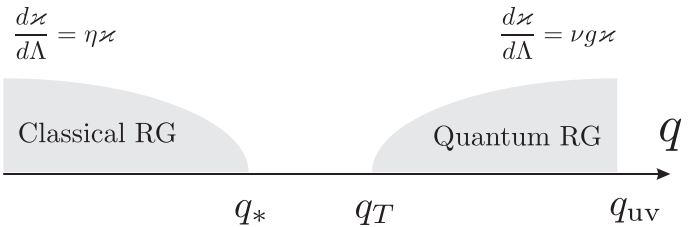} }
\caption{Characteristic momentum scales in the problem. Regions of quantum and classical RG are shown, along with the corresponding RG equations for bending rigidity. In these equations, $\eta$ and $\theta$ are  the  classical and quantum critical indices, respectively, while $g$ is the quantum coupling constant.
    }
\label{Fig2}
\end{figure}

In order to complete the calculation of $\alpha_T$, we should
find renormalization of $\varkappa$ in the whole interval  of momenta, $0< q <q_{\rm uv}$, and substitute the renormalized function $\varkappa_q$ into Eqs.~\eqref{xi}-\eqref{omega-q}.    This will allow us to determine the global stretching factor as a function of the applied stress and temperature, $\xi=\xi(\sigma , T),$  and thus to evaluate the thermal expansion coefficient,
\be
\label{alpha-T}
\alpha_T= \left[\frac{\p \xi^2(\sigma , T)}{\p T}\right]_\sigma.
\ee
 Hence, we focus in the  next subsection on renormalization  of  elastic constants.

 \subsection{Renormalization group}
 \label{sec-RG}

 In this section,  we find  the   flow  of the  elastic moduli with decreasing $q$ down from $q_{\rm uv}$ by using the perturbative RG approach (for a recent discussion of a quantum non-perturbative RG  see  Ref.~\onlinecite{moh}).  The classical and quantum RG are separated by $q=q_T$  found from
the condition $\hbar \omega_q \simeq T $:
\be
q_T \sim \frac{q_*}{\sqrt g}.
\label{qT}
\ee
For $g\ll 1,$  the characteristic scales of the problem are shown in Fig.~2.    At large spatial scales (for $q\ll q_*$),  a classical RG apply, so that the bending rigidity scales according to Eq.~\eqref{kappa-q}.   In the interval  $  q_* <q < q_T,$   the FP frequency is  still  small compared to $T$, so that  classical  approach  remains applicable. However, the renormalization of $\varkappa$ in this region is small  \cite{Nelson}, $ (\varkappa_q - \varkappa) /\varkappa   \sim   q_*^2/q^2$.  The quantum RG operates in the interval between  $q_T$ and $q_{\rm uv}$, with the former scale serving as an infrared and the latter as an ultraviolet cutoff.
In order to derive quantum RG equations, we notice that for $q<\sqrt{\mu_0/\varkappa_0},$
  the FP are much softer than the in-plane modes:
  \be \omega_q \ll \omega_q ^{\perp,  \parallel},
  \label{cond1}
  \ee
  where
  $$\omega_q ^{\perp} =q \sqrt{\mu/\rho}, \qquad \omega_q ^{\parallel}=q \sqrt{(2\mu+\lambda)/\rho}$$
  are frequencies of the transverse and longitudinal in-plane phonos respectively. We further notice, that
   for graphene, the value of  $ \sqrt{\mu_0/\varkappa_0}$ is on the order of the inverse lattice constant.
This implies that one can use this value as   the ultraviolet  atomic momentum scale:
\be q_{\rm uv} \sim \sqrt{\mu_0/\varkappa_0}. \label{q-uv}\ee
   In view of Eq.~\eqref{q-uv}, the  interval $q_T<q<q_{\rm uv}$  for quantum RG exists  provided that the temperature is not too high, $T < T_{\rm uv}$, where
\be
T_{\rm uv} \sim  g_0\varkappa_0.
\label{condition-T}
\ee
For higher temperatures one can fully neglect quantum effects. Equation \eqref{condition-T} implies that for graphene ($g_0\simeq 1/20$, $\varkappa_0\simeq 1$ eV) the temperature  $T_{\rm uv}$ is of the order of 500 K. This temperature plays the role of the Debye temperature in our model.

As follows from Eq.~\eqref{cond1}, the retardation effects can be neglected and  the quantum  RG  flow can be found in a full analogy  with classical RG,  \cite{my-crump}  where FP field is considered to be static.  Technical details are described in  Appendices \ref{ren} and \ref{app-D} where two alternative derivations of RG equations are presented.   For in-plane moduli, one gets the following RG equations:
\BEA
\label{RG-Y}
\frac{d}{d\Lambda} \frac{1}{Y}&=& \frac{3}{8}\frac{d }{d\Lambda} \frac{1}{ B}  =
\frac{d_c g}{(6+d_{\rm c}) Y},
\EEA
where
$$ \Lambda=\ln\left( {q_{\rm uv}}/{q}\right)$$
is the logarithm of the running RG scale $q$,
and the coupling $g$ is given by Eq.~\eqref{g0}.  For $d_{\rm c}=1,$ these equations are equivalent to Eqs.~(9) and (10) of Ref.~\onlinecite{kats2}.  As seen, there is invariant subspace of elastic moduli,  $Y=8B/3,$  which is conserved by the  RG flow.

 The RG flow of  the bending rigidity $\varkappa$ is given by equation
\be
\frac{d\varkappa}{d \Lambda}= \frac {4 g \varkappa} {d_c+6}   ,
\label{kappa0}
\ee
which agrees up to a sign with Eq.~(11) of Ref.~\onlinecite{kats2}. [The sign was recently corrected in Ref.~\onlinecite{kats-erratum}.]
From Eqs.~\eqref{g0}, \eqref{RG-Y}, and  \eqref{kappa0}, we find
\be
\frac{dg}{d \Lambda}= -g^2,
\label{Rg-g}
\ee
which again agrees with the result of Ref.~\onlinecite{kats2} once the sign error is corrected \cite{kats-erratum}.
 (We notice that the numerical coefficient in definition of $g$  in  Ref.~\onlinecite{kats2} is different).
 The negative sign in Eq.~(\ref{Rg-g}) is of key importance: it implies that the quantum anharmonicity effects {\it stabilize} the membrane increasing the effective bending rigidity. In other words, the flat phase of the membrane is perfectly defined in the limit of zero temperature and infinite system size. This is crucially important
 for low-temperature behavior of  the $\alpha_T$
 It is the growing effective bending rigidity at low temperatures which suppresses the thermal expansion coefficient. However, contrary to all previously known situations this provided only logarithm-in-power vanishing of the thermal expansion coefficient at $T \rightarrow 0$ rather than power-law.

 Solving the RG equations, we get the following flow of the quantum coupling constant $g$, the bending rigidity $\kappa$, and the in-plane moduli $Y$ and $B$:
  \BEA
\hspace*{-1cm}&&g=\frac{g_0}{1+g_0\Lambda}, \qquad \varkappa= \varkappa_0 (1+g_0\Lambda)^{\theta},
\label{RG-g-kappa0}
\\
\hspace*{-1cm}&&Y\!=\!\frac{Y_0}{(1+g_0\Lambda)^{1-3\theta/2}},  \qquad \frac{1}{B}\!=\! \frac{1}{B_0}\!+\!\frac{8}{3}\! \left(\frac{1}{Y}\! -\!\frac{1}{Y_0}\right).
\label{RG-Y-N}
\EEA
Here
 \be
 \theta=\frac{4}{d_c+6}
 \ee
 is the  quantum anomalous exponent. For graphene (or, more generally, for a 2D membrane in a 3D space), $d_c=1$   and $\theta=4/7.$    The bare coupling constant $g_0$ is quite small for graphene (about 1/20). With increasing spatial scale,  the running coupling constant   $g$  decreases according to RG equation \eqref{RG-g-kappa0} from $g_0$ (equal to $\simeq$ 1/20 for graphene) down to  zero. Hence, at all scales $g\ll1.$  This justifies \cite{kats2} the applicability of  one-loop RG approach.

The quantum  RG   stops     at $q=q_T$.  The overall picture of the renormalization of $\varkappa$ is as follows. The  RG flow starts at $q\sim q_{\rm uv},$ where  $\varkappa$ has a bare value $\varkappa_0.$   In the interval  $  q_T<q <q_{\rm uv},  $   the bending rigidity grows according to Eq.~\eqref{RG-g-kappa0}.  The value of $\varkappa_q$ at the edge of the quantum interval (at $q \simeq q_T$) and the corresponding value of $g$ are given by
$$
\varkappa=\varkappa (\Lambda_T), \qquad g=g(\Lambda_T),
$$
where  $\varkappa (\Lambda)$ and $g=g(\Lambda)$ are given by Eq.~\eqref{RG-g-kappa0}
and
$$
\Lambda_T =\ln\left( {q_{\rm uv}}/{q_T}\right) \approx \ln \sqrt{T_{\rm uv}/T.}
$$
Below the couplings without indication of momentum scale (such as $g$, $\varkappa$, $B$, $Y$) will be understood as defined on the scale $\Lambda_T$ governed by the temperature.
It is worth noting that values of  $q_*$ and $q_T$ are determined by the renormalized elastic  moduli:
  \BEA
  && q_T \sim \frac{\sqrt{Y T}}{\varkappa \sqrt{g}} \sim  \frac{\sqrt{Y_0 T}}{\varkappa_0}\frac{1}{ \sqrt{ g_0} (1+g_0\Lambda_T)^{\theta/4} } ; \\
  && q_* \sim \frac{\sqrt{Y T}}{\varkappa} \sim  \frac{\sqrt{Y_0 T}}{\varkappa_0}\frac{1}{ (1+g_0\Lambda_T)^{(2+\theta)/4} }.
  \EEA
 In the interval    $q_*<q<q_T$, the bending rigidity does not change essentially (i.e., it changes by a factor of order unity). Finally,  at  lowest  momenta   $q <q_*, $  the bending rigidity scales according to Eq.~\eqref{kappa-q} with $\varkappa =\varkappa (\Lambda_T).$

Now we are in a position to calculate the integrals entering Eqs.~\eqref{xi0} and \eqref{S} and to get final formulas governing the thermodynamics of graphene.   Results of this analysis are presented in the next section.

 \section{Results}
 \label{sec.III}

 \subsection{Thermal expansion coefficient: Zero tension}

 The  main contribution to  $\xi_0$  as given by Eq.(\ref{xi0}) comes from the region $q<q_T.$  For $\eta \ll1, $  a simple calculation yields
      \be
      \xi_0^2\approx 1-\frac{d_{\rm c}T}{8\pi\varkappa}\left[\frac{2}{\eta}+ \ln\left( \frac{1}{g}\right) \right].
      \ee
Here we neglect terms of the order of $g_0$ as well as  ones of the order of $T/\varkappa$ coming from $q>q_T.$ We, thus, obtain the thermal expansion coefficient of an unstressed membrane:
\be
\label{alphaT-q}
 \alpha_T\approx -\frac{d_{\rm c}}{8\pi  \varkappa}\left[ \frac{2}{\eta}+ \ln\left( \frac{1}{{ g}}\right) \right].
 \ee
Two terms in the square brackets  represent contributions of momentum intervals $q<q_*$ and $q_*< q <q_T$,  respectively. Comparing  Eq.~\eqref{alphaT-q} with Eq.~\eqref{alphaT-d-infty}, we  observe two differences.

Firstly,  a logarithmic-in-$g$ term (reflecting the contribution of the momenta $q_*< q <q_T$) appeared  in the square brackets.  This term can be neglected   for a generic   membrane embedded in the space of high dimensionality ($d_{\rm c} \to \infty, \eta \to 0$)  \cite{my-hooke}. However, for graphene, where $\eta \simeq 0.8$ and $1/g_0 \simeq 20$, the two terms give comparable contributions.  On the other hand, for a ``nearly classical'' 2D membrane in 3D space that has the same $\eta$ and much larger $1/g_0$, the logarithmic contribution would be dominant.

Secondly,  $\alpha_T$ becomes now a slow function of temperature. This dependence deserves a special attention.
 As follows from Eqs.~\eqref{RG-g-kappa0} and \eqref{alphaT-q},
  $\alpha_T$  remains negative and nearly constant in an extremely wide  temperature range:
\be
\alpha_T\approx \alpha_{\rm max}
=  \!- \frac{d_c}{8\pi  \varkappa_0}\left[ \frac{2}{\eta}+ \ln\left(\! \frac{1}{{g_0}}\!\right) \right], \quad T_0 \ll T  \ll T_{\rm uv} .
\label{alpha-T-constant}
\ee
Here  $T_0 $   is  given by Eq.~\eqref{Tuv}, yielding for graphene   $T_0 \simeq 10^{-14}$ K. Thus the thermal expansion coefficient of graphene remains nearly constant within almost twenty decades of temperature! Using graphene parameters ($d_{\rm c}=1,$ $\varkappa \simeq 1$ eV), we get an estimate for this constant value:
$\alpha_T \simeq- 0.23\:{\rm eV}^{-1}$.

\begin{figure}[t]
\centerline{\includegraphics[width=0.5\textwidth]{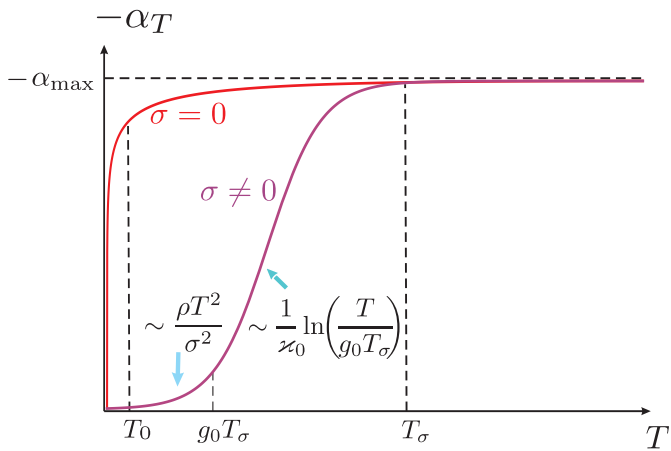} }
\caption{Temperature dependence of the thermal expansion coefficient for unstressed ($\sigma =0$) and stressed ($\sigma \ne 0$) membrane. The tension $\sigma $  suppresses the thermal expansion  at  $T<T_\sigma$.
}
\label{Fig3}
\end{figure}

Only at exponentially low temperatures, $T\ll T_0$, the thermal expansion coefficient starts to decay logarithmically with decreasing temperature:
\be
\label{alphaT2-q}
\alpha_T \simeq -\frac{d_c}{8\pi \varkappa_0 } \frac{\ln \ln (T_{\rm uv}/T) }{[(g_0/2) \ln (T_{\rm uv}/T) ]^ \theta}, \qquad T\ll T_0.
\ee
The temperature dependence of $\alpha_T$  is shown in Fig.~\ref{Fig3} for $T\ll T_{\rm uv}$.

It is instructive to analyze how the classical limit ($\hbar \to 0$, implying $g_0\to 0$), is approached. We recall  that the value of $\alpha_T$ depends logarithmically on $g_0$. Consequently, the thermal expansion coefficient (\ref{alpha-T-constant}) diverges in the classical limit. It should be emphasized, however, that the range of validity of Eq.~(\ref{alpha-T-constant}) shrinks in this limit in view of Eq.~(\ref{condition-T}). For temperatures above $T_{\rm uv}$, the logarithmic term in Eq.~(\ref{alpha-T-constant}) gets modified, becoming temperature-dependent:
\begin{equation}
\alpha_T\approx - \frac{d_c}{4\pi  \varkappa_0}\left[ \frac{1}{\eta}+ \ln\left( \frac{q_{\rm uv}}{{q_*}}\right) \right], \qquad  T \gg  T_{\rm uv} .
\label{alpha-T-high-temp}
\end{equation}
In other words,  the function $|\alpha_T (T)|$  has a maximum  at $T\simeq T_{\rm uv}$ (this maximum is not shown in Fig.~3, which is plotted  for $T\ll T_{\rm uv}$). That is, it is  the maximal value  $\alpha_{\rm max}$  that diverges in the classical limit: $\alpha_{\rm max} \propto -\ln (1/g_0).$

 \subsection{Thermal expansion coefficient:  Finite tension}

We turn now to a generalization of the above results to the case of non-zero tension,  $\sigma \neq 0.$  In this case,
  there appears a new characteristic momentum  $q_\sigma$ determined by the condition
  $$\varkappa_q q_\sigma^2 \sim \sigma.$$
  This momentum  increases with $\sigma $, reaching the value of $q_*$    for
  $\sigma =\sigma_*$,   where
   \be
   \sigma_*   \sim  \varkappa_{q_*} q_*^2   \sim  \frac{d_{\rm c}Y T}{ \varkappa} .
   \label{sigma-star}
   \ee
We remind the reader that, for temperatures below $T_{\rm uv}$, couplings without the momentum indicated [such as $Y$ and $\varkappa$ in Eq.~(\ref{sigma-star})] are understood as those including the quantum renormalization, i.e., defined on the temperature scale $q_T$.   We have also taken into account in Eq.~(\ref{sigma-star}) that there is no essential renormalization of $\varkappa$ between the scales $q_T$ and $q_*$.

The  physical meaning of $\sigma_*$ was discussed  in Ref.~\onlinecite{my-hooke}.  For $\sigma>\sigma_*$ the  membrane shows linear  Hooke's law,  \cite{my-hooke}, while for $\sigma<\sigma_*$  the stress-strain relation is of  a power-law form,
$$
\xi-\xi_0 \propto \sigma^\alpha,
$$
 with an anomalous exponent $\alpha$   given by
 $$
 \alpha=\frac{\eta}{2-\eta}.
 $$
One can express $\sigma_*$ in terms of the bulk modulus
   \be
   \sigma_*= \frac{d_{\rm c} C B T}{ 4\pi \varkappa},
   \label{CC}
   \ee
   where $C$ is a numerical coefficient of order unity, which is chosen from the requirement that the low-stress
   deformation takes the form given by upper line of Eq. (\ref{scl1}) below. This coefficient is determined by relations between  values of elastic constants.   (In principle, $C$ depends on the ratio of elastic module and, therefore, is a slow function of temperature    \cite{comment}.)  The room-temperature value of $C$ for graphene, $C\simeq 0.25$, was found  in Ref.~\onlinecite{my-hooke} from a comparison of the analytic strain-stress relation with results of atomistic simulations of Ref.~\onlinecite{katsnelson16}. [The definition of the coefficient $C$ in Eq.~(\ref{CC}) differs from that in Ref.~\onlinecite{my-hooke} by an additional factor $2/\eta,$ which is   $ \simeq 3$ for physical membranes.]


    With further increase of $\sigma$ up to the value $\sigma_*/g, $   the momentum $q_\sigma$ reaches the boundary of the quantum region:
    \be
    q_\sigma \simeq q_T \quad \text{for} \quad \sigma  \simeq \sigma_*/g.
    \ee
Since we want to analyze a temperature dependence of membrane  properties  at non-zero tension $\sigma$, it is useful to introduce a characteristic temperature  $T_\sigma$ determined by the condition $\sigma=\sigma_*(T)$,
   \be
   T_\sigma =\frac{4\pi  \varkappa \sigma}{d_{\rm c}CB} \sim \varkappa \frac{\sigma}{ B } .
   \ee

 The stress-induced deformation $s(\sigma)$ can be separated into classical and quantum parts:
 $$
 s(\sigma)=s_{\rm cl} (\sigma)+s_{\rm quant}(\sigma),
 $$
 where
\BEA
&&s_{\rm cl} \approx \frac{d_{\rm c} T}{4\pi \rho}\int\limits _0^{q_T} dq q^3 \left(\frac{1}{(\omega_q^0)^2} -\frac{1}{\omega_q^2 } \right),
\label{scl}
\\
&& s_{\rm quant} \approx \frac{d_{\rm c} }{8\pi \rho}\int\limits ^{q_{\rm uv}}_{q_T} dq q^3 \left(\frac{1}{\omega_q^0} -\frac{1}{\omega_q} \right).
\label{sqw}
\EEA
For $\sigma <\sigma_*/g$, the classical contribution is essentially non-perturbative with respect to $\sigma$ (see Ref.~\onlinecite{my-hooke}):
\be \label{scl1}
s_{\rm cl} \approx \frac{\sigma_*}{ B} \left \{ \begin{array}{ll}
                        (1/\alpha)\left(\sigma/\sigma_*\right)^\alpha,  &~ \text{for}~ \sigma<\sigma_* \\[0.2cm]
                        \displaystyle
                       ({1 }/{2 C})[2/\eta+\ln(\sigma/\sigma_*)], &
                     ~\text{for}~ \sigma_*<\sigma<\sigma_*/g.
                     \end{array}
 \right.
\ee
  In  contrast,  the quantum contribution can be calculated perturbatively by expansion to the leading order in $\sigma$:
\BEA
\nonumber
s_{\rm quant} & = & - \frac{d_{\rm c} \sigma}{8\pi\rho} \int \limits_{q_T}^{q_{\rm uv}}dq q^3 \left(\!\frac{\p \omega_q^{-1} }{\p\sigma}\!\right)_{\!\sigma=0} \\
&=& \sigma \int \limits_0^{ \Lambda_T} d\Lambda \frac{\p  B^{-1} }{\p \Lambda}= \frac{\sigma }{B}-\frac{\sigma}{B_0}.
\EEA
Here we have used Eq.~\eqref{RG-Y}. Substituting $s_{\rm quant}$ into Eq.~\eqref{xi}, we see that quantum effects lead  to  a simple renormalization of the  first term in the r.h.s. of Eq.~\eqref{xi}: $\sigma/B_0 \to \sigma/B.$

For  a sufficiently large stress (or for a sufficiently low temperature),  $\sigma > \sigma_*/g $, the momentum $q_\sigma$ becomes larger than $q_T$, so that one can neglect the term $\varkappa_q q^4$  in Eq.~(\ref{omega-q}) in comparison with $\sigma q^2$ in the whole classical region.  In this case, Eq.~\eqref{xi} takes the form
$$
\xi^2=\xi_\sigma^2+ \frac{\sigma}{B_\sigma},
$$
where   $B_\sigma$ is the renormalized value of the  bulk modulus [see Eq.~\eqref{RG-Y-N}] at the scale $q_\sigma \simeq \sqrt{\sigma/\varkappa},$ and
$$
\xi_\sigma^2\approx 1-\frac{d_{\rm c}}{8\pi \sqrt {\rho \sigma}} \int dq\, q^2
\left[\coth \left (\frac{\sqrt \sigma q}{ 2 \sqrt \rho T}\right)  - 1\right] .
$$
Here we neglected small ($\sim g$) terms.
The integral in this formula scales with  decreasing temperature as $T^3.$

Summarizing the obtained results, we find
 \be
\xi^2 -1\!\approx\! \left\{\!  \begin{array}{c}
                       \displaystyle -   \frac{d_{\rm c} T}{8\pi\varkappa} \left[  \frac{2}{\eta}+ \ln\left(\frac{1}{g}\right)\right] + \frac{\sigma_*}{\alpha B}  \left (\frac{\sigma}{\sigma_*}\right)^\alpha\!\!,
                             ~\text{for} ~ \sigma<\sigma_*,
                             \\[0.35cm]
                            \displaystyle \frac{\sigma}{ B} - \frac{d_c T}{8\pi \varkappa} \ln\left(\frac{\sigma_*}{\sigma g}\right),  ~ \text{for} ~ \sigma_*<\sigma<\sigma_*/g,
                             \\[0.35cm]
                            \displaystyle  \frac{\sigma}{ B_\sigma} - \frac{d_c \rho T^3  \zeta(3)}{2\pi \sigma^2},
                              ~ \text{for}~  \sigma_*/g<\sigma.
                                                       \end{array}
\right.
\label{final-xi}
\ee
From
  Eqs.~(\ref{alpha-T}) and \eqref{final-xi},
 we obtain   the thermal expansion coefficient $\alpha_T$ as a function  of applied  stress:
\be
\alpha_T \approx -\frac{d_{\rm c }}{8\pi\varkappa} \left\{  \begin{array}{c}
                             2/\eta+ \ln(1/g) -C_1 (\sigma/\sigma_*)^\alpha,
                             \quad \text{for} ~ \sigma<\sigma_*,
                             \\[0.2cm]
                             \ln(\sigma_*/\sigma g), \quad \text{for} ~ \sigma_*<\sigma<\sigma_*/g,
                             \\[0.2cm]
                              12\zeta(3) \rho T^2 \varkappa/\sigma^2, \quad \text{for}~  \sigma_*/g<\sigma,
                                                       \end{array}
\right.
\label{final-alpha}
\ee
where $C_1= 4C(1-\eta)/ \eta$ and $\zeta(3)\simeq 1.202$ is the Reimann zeta function.

 The dependence of $\alpha_T$ on $T$ for stressed and unstressed membrane following from Eq.~\eqref{final-alpha} is shown in Fig.~3.  (In this figure, we assume for the stressed case that $g_0T_\sigma \gg T_0 .$)
 At high temperatures, $T\gg T_\sigma$,   the external tension results in a power-law dependence of $\alpha_T$ on temperature:
\begin{equation}
\alpha_T \approx -\frac{d_{\rm c}}{8\pi\varkappa} \left [ \frac{2}{\eta}+\ln \frac{1}{g}  -C_1 \left (\frac{T_\sigma}{T}\right )^{\alpha}\right ] .
\end{equation}
Below $T_\sigma,$  the absolute value of the thermal expansion coefficient decreases logarithmically:
\begin{equation}
\alpha_T \approx -\frac{d_{\rm c}}{8\pi\varkappa}
                             \ln\frac{T}{ g T_\sigma}, \qquad  g T_\sigma  \ll T \ll T_\sigma.
\end{equation}
Finally, at still lower temperatures, $T\ll g  T_\sigma$,  we find  $\alpha_T \propto T^2$, see the third line in Eq.~\eqref{final-alpha}.

 \begin{figure}[t]
\centerline{\includegraphics[width=0.5\textwidth]{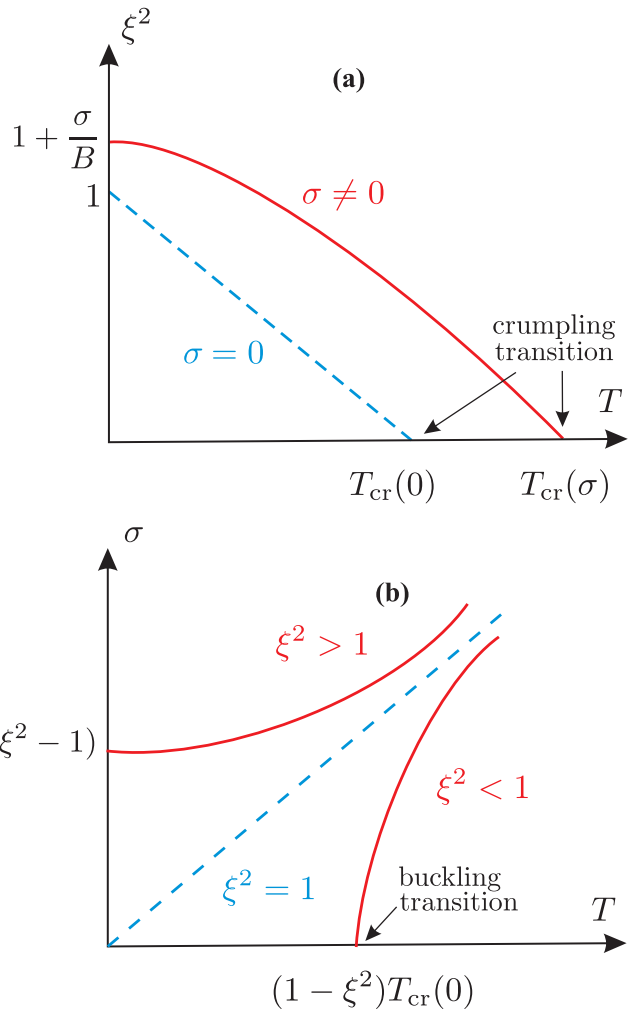} }
\caption{  Temperature dependences of (a) strain $\xi$ for fixed stress  $\sigma$, and (b) stress for fixed strain.   Positions of crumpling (in the left panel) and buckling (in the right panel, for $\xi <1$) transitions are indicated. }
\label{Fig4}
\end{figure}
\subsection{Effective bulk modulus}
 Differentiating Eq.~\eqref{final-xi} over $\sigma,$ we  find expression for effective bulk modulus
\BEA
&&\frac{1}{B_{\rm eff}}=\left(\frac{\p\xi^2}{\p \sigma} \right)_T \label{Beff}
\\[0.2cm]  \nonumber
&&=\left\{  \begin{array}{c}
                             (1/B) (\sigma_*/\sigma)^{1-\alpha},
                             \quad \text{for} \quad \sigma<\sigma_*,
                             \\[0.2cm]
                              1/B +  d_{\rm c} T/8\pi \varkappa \sigma, \quad \text{for} \quad \sigma_*<\sigma<\sigma_*/g,
                             \\[0.2cm]
                             1/B_\sigma +d_{\rm c} \zeta(3) \rho T^3 /\pi \sigma^3, \quad \text{for} \quad \sigma_*/g<\sigma,
                                                       \end{array}
\right.
\EEA
Two upper lines of Eq.~\eqref{Beff} were obtained previously in Ref.~\onlinecite{my-hooke}. In the second and third line we took into account small temperature-dependent corrections.  The third line shows that $B_{\rm eff}$ slowly increases with $\sigma$ at $\sigma > \sigma_*/g$ due to suppression of quantum RG by external stress:
\be
B_{\rm eff} \simeq B_\sigma =  \frac{B_0}{[1 +(g_0/2) \ln(\mu_0/\sigma)]^{1-3\theta/2}}.
\ee
[Here we  assume for simplicity that $Y=8B/3,$ use Eq.~\eqref{q-uv} and write $\ln(q_{\rm uv}/q_\sigma)$ with the logarithmic precision].
\subsection{General phase diagram  of membrane }
In Fig.~\ref{Fig4}, we plot the temperature dependence of the strain $\xi$ for fixed stress $\sigma $ (Fig.~\ref{Fig4}a) as well as the temperature dependence of the stress for fixed strain (Fig.~\ref{Fig4}b).   As seen from   Fig.~\ref{Fig4}a, with increasing  temperature for the fixed $\sigma,$ the membrane undergoes crumpling transition ($\xi \to 0$).  Corresponding critical temperature  increases  with the  applied stress  \cite{my-hooke}.  The dependence $T_{\rm cr} (\sigma)$  can be  found from the upper line of Eq.~\eqref{final-xi} by requirement $\xi=0$ and by taking into account Eqs.~\eqref{CC} and \eqref{RG-g-kappa0}.
The  plot shown in Fig.~\ref{Fig4}b  corresponds to an experimental setup in which the membrane in-plane area is kept fixed, while the temperature is varied. If this area is smaller than the intrinsic zero-temperature area of the membrane (i.e., $\xi < 1$), the membrane udergoes in the process of cooling a buckling transition.

  \subsection{Specific heat}

We evaluate now the  temperature dependence of the specific heat. Similar to the thermal expansion coefficient $\alpha_T$,   both constant-volume ($C_V$) and  constant-pressure ($C_P$) specific heat capacities are determined by FP.   To evaluate them, we first determine the entropy of the membrane,
$$
S=-L^{-2} (\p F / \p T)_{\xi},
$$
where $F$ is the free energy given by Eq.~\eqref{Fsigma0}.    We find
 \BEA
 &&S=d_{\rm c} \int \frac{d^2\mathbf q}{ (2\pi)^2}\left[  (N_q+1)\ln(N_q+1) - N_q \ln N_q \right.
 \nonumber
 \\
 &&\left. - (N_q+1/2) (\p \omega_q /\p T)_\sigma \right],
 \label{SSS}
 \EEA
where $N_q=[\exp(\omega_q/T) -1]^{-1}$ is the Bose function.   It is worth noting that the last term in the square brackets in Eq.~\eqref{SSS} is nonzero because of the temperature-dependent renormalization of the bending rigidity $\varkappa$ [see Eq.~\eqref{kappa-q}].  By using Eq.~\eqref{SSS}, one can calculate the specific heat capacities,
$$
C_{P,V} =-T  \left(\p S/\p T \right)_{P,V}.
$$
A straightforward calculation shows that, in the leading order, both $C_{P}$  and $C_{V}$ are   given by the classical formula for phonons with a parabolic spectrum,
\be
C_{P}  \approx C_{V} \approx d_{\rm c} \int \frac{d^2\mathbf q}{ (2\pi)^2} \omega_q \frac{\p N_q}{\p T } \approx \frac{\pi d_{\rm c}\sqrt \rho T }{12 \sqrt \varkappa}.
\ee

In the presence of a finite tension, this result is valid for temperatures $T \gg g T_\sigma$ (or, equivalently, the tension $\sigma \ll \sigma_*/g$).  At low temperatures, $T\ll g T_\sigma$, the specific heat is proportional to $T^2$:
\begin{equation}
C_P\approx C_V \approx
\frac{ 3 \zeta(3) d_{\rm c}\rho}{\pi \sigma}  T^2.
\end{equation}
While the leading contribution to both $C_{P}  $ and $ C_{V}$  is due to   classical fluctuations, the difference $C_{P}  - C_{V}$ is small, proportional to $g,$ and, therefore, is due to the quantum effects:
\BEA
\hspace*{-1cm}
&&\frac{C_P-C_V}{C_P} = T\frac{(\p \xi^2/\p T)^2_\sigma}{ C_P (\p \xi^2/\p \sigma)_T}
 \\[0.2cm]\nonumber
 & \sim&  g  \left\{  \begin{array}{ll}
                    \ln^2\left( {1}/{g}\right) (\sigma/\sigma_*)^{1-\alpha}, &\ \ \text{for} ~  \sigma\ll \sigma_*, \\[0.15cm]
                     \left [\ln(\sigma_*/g\sigma)\right]^2, & \ \ \text{for} ~  \sigma_* \ll \sigma\ll \sigma_*/g,\\[0.15cm]
                     (B_\sigma/B) (\sigma_*/g\sigma)^3,& \ \ \text{for} ~  \sigma_* /g\ll \sigma
                 \end{array}
    \right.
\EEA
Hence, at high temperatures, $T\gg T_\sigma$, we find
\begin{equation}
\frac{C_P-C_V}{C_P}  \sim g \ln^2\left( \frac{1}{g}\right)   \left ( \frac{T_\sigma}{T}\right )^{1-\alpha}.
\label{cpcv1}
\end{equation}
At intermediate temperatures, $g T_\sigma \ll T \ll T_\sigma$, the result reads
\begin{equation}
\frac{C_P-C_V}{C_P}  \sim g  \ln^2 \frac{T}{g  T_\sigma} ,
\label{cpcv2}
\end{equation}
while at very law temperature $T \ll g T_\sigma ,$ we obtain
\begin{equation}
\frac{C_P-C_V}{C_P}  \sim
g   \left(\frac{T}{g  T_\sigma}\right)^3 \frac{B_\sigma}{B} \sim \frac{\rho T^3 B_\sigma}{ \sigma^3}
.
\label{cpcv3}
\end{equation}
We emphasize that
$C_P-C_V$ vanishes in the absence of the external tension ($\sigma=0$), as follows from Eq.~(\ref{cpcv1}) with $T_\sigma =0$.

\subsection{Gr\"uneisen  parameter}
Finally, we consider the macroscopic Gr\"uneisen  parameter
\be
 \gamma=\frac{ \alpha_T(\p \sigma/\p \xi^2)_T}{C_V},
 \ee
which is an important characteristics of thermomechanical properties of a system.
 We find that the Gr\"uneisen  parameter is negative for all values of stress:
 \be
\gamma \sim - \frac {g \varkappa}{T} \left \{  \begin{array}{cc} \displaystyle
  \ln(1/g) \left({\sigma}/{\sigma_*}\right)^{1-\alpha}, &  ~\text{for} ~  \sigma \ll \sigma_*, \\[0.4cm]
                      \ln(\sigma_*/g \sigma), & ~ \text{for} ~  \sigma_* \ll  \sigma\ll\sigma_*/g,
                      \\[0.4cm]
                      (B_\sigma/B) (\sigma_*/g\sigma),& ~ \text{for} ~  \sigma_* /g\ll  \sigma.
                 \end{array}
   \right.
 \ee
 Several points deserves special attention. First of all, we see that the absolute value of $\gamma$ has a maximum as a function of $\sigma$ at $\sigma \simeq \sigma_*.$   On the other hand, for fixed $\sigma,$  $|\gamma|$ is a monotonously decreasing function of $T.$ Most importantly,  in the limit of low temperature, $\gamma $ turns out to be non-zero
 \be
- \gamma   \sim  \frac{B_\sigma}{\sigma}, \quad \text{for}~T\to0.
 \ee
  Schematic  dependence of   the Gr\"uneisen  parameter on tension and  temperature  is illustrated in Fig.~5.

\begin{figure}[t]
\centerline{\includegraphics[width=0.45\textwidth]{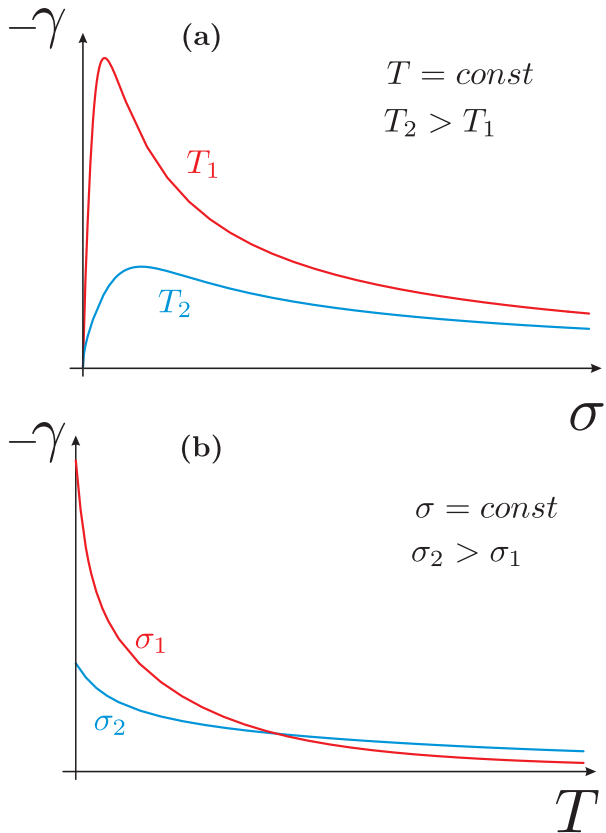} }
\caption{ Schematic  dependence of   the Gr\"uneisen  parameter on tension at fixed temperature (a) and on temperature  for fixed  tension (b). }
\label{Fig5}
\end{figure}

\subsection{Third law of thermodynamics}
 Let us point out that the  third law of thermodynamics manifests itself in this problem in  a somewhat curious way.  Indeed, the entropy  \eqref{SSS} vanishes in the limit $T \to 0$  for any $\sigma$  even if the quantum renormalization effects are neglected. On the other hand, it is easy to see   from the first line of Eq.~\eqref{final-alpha} that  $\alpha_T|_{\sigma=0}$ remains finite in the limit $T \to 0.$ At first glance, this may seem to contradict to Eq.~\eqref{third} that yields
$\alpha_T=\p S/\p \sigma.$  This apparent contradiction is resolved by noticing that non-analyticity of $S$  at $T= 0, ~\sigma = 0$  leads to non-commutativity of limits $T \to 0$ and $\sigma \to 0$   for the function $\alpha_T(T,\sigma).$  Quantum renormalization effects restore the vanishing of the thermal expansion coefficient at  $T = 0$  independently of the order of  the limits, which is conventionally considered as a manifestation of the third law.

\section{Conclusion \label{con}}

To summarize, we have developed a theory of thermomechanical properties of a suspended graphene membrane.
We have shown that at zero tension the thermal expansion coefficient  $\alpha_T$ of    free-standing graphene   is negative and  temperature-independent  in a very broad temperature range, see Eq.~(\ref{alpha-T-constant}).  The underlying physics of  the negative expansion is  the global shrinking of the graphene  membrane  in the longitudinal direction due to classical transverse fluctuations.

The second term in Eq.~(\ref{alpha-T-constant})  for $\alpha_T$ is governed by the  dimensionless quantum coupling constant, $g_0 \ll 1$. This coupling constant vanishes in the classical limit $\hbar \to 0$ (thus implying a divergence of $\alpha_T$) and is equal to $\simeq 0.05$ for graphene. The small value of $g_0$ ensures that $\alpha_T$ remains $T$-independent down to extremely low temperature $T_0$, Eq.~(\ref{Tuv}).
For graphene parameters, we estimate the value of the thermal expansion coefficient as  $\alpha_T \simeq  - 0.23\:{\rm eV}^{-1}$,   which applies below the temperature $T_{\rm uv} \sim g_0 \varkappa_0 \sim 500$\:K  (where $\varkappa_0 \sim 1$\:eV is the bending rigidity) down to $T_0 \sim 10^{-14}$\:K.
For $T<T_0$, the absolute value of the thermal expansion coefficient starts to decrease logarithmically slowly with decreasing temperature, Eq.~(\ref{alphaT2-q}), since quantum effects lead to increase of the bending rigidity.
Our results  imply that, contrary to naive expectations, quantum fluctuations do not lead to to the melting or crumpling of a  2D crystal but instead stabilize the membrane due to enhanced role of the anharmonicity.

A finite tension $\sigma $  suppresses the thermal expansion  at  $T<T_\sigma$, where $T_\sigma \propto \sigma $ is the characteristic temperature which separates regimes of  conventional ($T<T_\sigma$) and anomalous ($T>T_\sigma$) elasticity,  see Fig.~\ref{Fig3}.

We  have also evaluated the temperature dependence of tension in a  graphene membrane placed into a frame of a fixed size $\xi L$. With lowering temperature, a membrane with $\xi<1$ undergoes then a  buckling transition,  see Fig.~\ref{Fig4}b.

Finally, we have  calculated the specific heat of the membrane. We have found that in the leading order both  $C_P$ and $C_V$ are dominated by classical effects and are  given by a standard expression for phonons with parabolic spectrum. On the other hand,   a small difference $C_P-C_V$  is due to quantum fluctuations and shows a very non-trivial behavior as a function of  the ratio $T/T_\sigma$, see Eqs.~(\ref{cpcv1}), (\ref{cpcv2}), and (\ref{cpcv3}).  The same ratio $T/T_\sigma$ determines the temperature and stress dependence of the Gr\"uneisen parameter, which turns out to be negative for all temperatures and tensions, being monotonous function of  $T$ (for fixed $\sigma$) and showing a minimum  as a function of $\sigma $ (for fixed $T$)  for $\sigma \simeq \sigma_*.$

Our results demonstrate that 2D materials are dramatically different from 3D ones where, according to Gr\"uneisen law, the thermal expansion coefficient is proportoinal to the heat capacity and goes to zero as $T \to 0$ as a power law of temperature. It would be very interesting to check our prediction for the low-temperature behavior of $\alpha_T$ experimentally. In particular, the thermal expansion coefficient can be measured from the temperature shift of Raman spectra  \cite{Yoon}. An alternative (and possibly an easier) way of determining $\alpha_T$ is provided by studies of van der Waals heterostructures such as graphene/hBN, graphene/MoS$_2$, etc.  \cite{VdWGeim,VdWNovo}. In this setting, large thermal expansions of 2D materials at low temperatures would result in a strong temperature dependence of lattice mismatch which can be seen via reconstruction of Moire patterns  \cite{moire1,moire2,moire3} or via characteristics of graphene bubbles on a substrate  \cite{bubble}.

\section{Acknowledgement \label{ack}}

We thank  E. I. Kats, V. V. Lebedev, and K. S. Novoselov for useful discussions.
The work was supported by the Russian Science Foundation (grant No. 14-42-00044). MIK acknowledges a support by NWO via Spinoza Prize.

\appendix

\section{Calculation of free energy}\label{F-S}

In this Appendix we calculate the free energy corresponding to the Lagrangian \eqref{L}.
The  partition function reads
\be
Z=\int \{D\mathbf r\}e^{-S},\,\,\, S=\int_0^{\beta}d\tau \int d^2\mathbf x \:{\cal L},
\ee
where $\beta=1/T.$
In terms of $\mathbf u,\ \mathbf h$, and $\xi$ introduced in Eq.~(\ref{r}), the
 Lagrangian density becomes
 \be
 {\cal L}\! =\!\frac{\rho}{2}( \dot{\mathbf u}^2\!+\!\dot{\mathbf h}^2)\!+\!\frac{\mu\!+\!\lambda}{2}\! \left[ \!\left (\!\xi^2\!-\!1\!+\!\frac{K}{2}\! \right)^2\! -\!\frac{K^2}{4}\!\right]\!+\!{\cal L}_0,
 \label{Luh}
 \ee
 where $\dot{\mathbf u} =\p \mathbf u / \p \tau$, $\dot{\mathbf h} =\p \mathbf h / \p \tau,$
 \be
K=2 \left \langle  u_{\alpha\alpha} \right \rangle_{\mathbf x,\tau}= \left \langle  \p_\alpha \mathbf h  \p_\alpha \mathbf h+\p_\alpha \mathbf u  \p_\alpha \mathbf u \right \rangle_{\mathbf x,\tau},
\label{K}
\ee
 \be
  u_{\alpha\beta}= \big[\xi(\p_\alpha u_\beta+\p_\beta u_\alpha) +\p_{\alpha}{\mathbf h}\p_\beta{\mathbf h} +\p_{\alpha}{\mathbf u}\p_\beta{\mathbf u}\big]/2,
\label{strain}
 \ee
$\langle \cdots \rangle_{\mathbf x,\tau} $  denotes  the spatial and time averaging,
$$
\langle \cdots \rangle_{\mathbf x,\tau} = \int \frac{d^2\mathbf x}{L^2}\int_0^\beta \frac{ d\tau}{\beta}\cdots,
$$
and
\be
 {\cal L}_0=\frac{\varkappa}{2} [(\Delta\mathbf h )^2 +(\Delta\mathbf u)^2 ]+\mu  u_{ij}^2+\frac{\lambda}{2} u_{ii}^2.
\label{L0}
\ee
Equation \eqref{strain}  coincides with the conventional expression for the  strain tensor of the membrane provided that $\xi=1$ and  the term $\p_{\alpha}{\mathbf u}\p_\beta{\mathbf u}$ in the square brackets (which is of the second order in  $\p_{\alpha}{\mathbf u}$ and thus much smaller than the first term) is neglected.   Within such an approximation, and neglecting also a
 small term $\varkappa (\Delta\mathbf u )^2/2$ in Eq.~\eqref{L0},
the Lagrangian ${\cal L}_0 (\mathbf u,\mathbf h) $  coincides with the textbook expression for elastic energy of a nearly flat membrane. \cite{Nelson}

The term $(\mu+\lambda)K^2/8$ in Eq.~\eqref{Luh}  represents a quatric interaction (with $h^4$, $u^4$, and $h^2u^2$ couplings) with zero
transferred momentum and energy, $q=0$ and $\Omega=0$ (zero mode). After combining this term with the  analogous  quartic zero-mode  term  coming from ${\cal L}_0$,     the zero-mode contribution  can be safely neglected because it gives a  negligibly small correction  to the self-energies of both flexural and in-plain phonons (see detailed discussion in the Supplementary Material of Ref.~\onlinecite{my-hooke})

In the $(\mathbf q,\omega)$  representation, the action we are left with  reads
 \BEA
 \noindent
 S&=&
 \int \limits _{\mathbf q\omega}
  \left[ \frac{\rho\omega^2\! +\! \varkappa q^4}{2} (|\mathbf h^{\mathbf q,\omega}|^2\!
+\! |\mathbf u^{\mathbf q,\omega}|^2 \!) \right.
 \\ \nonumber
 &+& \left.
\mu | u_{\alpha \beta}^{\mathbf q,\omega}|^2 \!+\!
 \frac{\lambda}{2} | u_{\alpha \alpha}^{\mathbf q,\omega}|^2 \right]\! +\! \frac{\mu\!+\!\lambda}{2} \frac{L^2}{T}\left (\xi^2\!-\!1\!+\!\frac{K}{2}\!\right)^2.
  \label{Sqw}
  \EEA
Here  $\int \limits _{\mathbf q\omega} $ stands for $T\sum\limits_\omega \int d^2\mathbf q/(2\pi)^2 $
and summation goes over Matsubara frequencies  $\omega =2\pi n T.$      The next step is to integrate $\exp(-S) $ over  $\mathbf u$ and $\mathbf h.$  To carry out this integration, we  first perform a Hubbard-Stratonovich decoupling of the last term   in Eq.~(\ref{Sqw}) by
 an integral  over an auxiliary field $\chi$,
  \BEA
 && \exp \left[  -\frac{L^2(\mu+\lambda)}{2T}\left(\xi^2 -1+\frac{K}{2}\right)^2 \right] = \frac{L}{\sqrt{2\pi(\mu+\lambda) T}}
  \nonumber
 \\
 \nonumber
 && \times \int d\chi \exp\left\{-\left[\frac{\chi^2 }{2(\mu+\lambda)} - i \chi \left(\xi^2-1+\frac{K}{2}\right)\right]\frac{L^2}{T}\right\}.
 \EEA
This yields
\BEA
e^{-S}&=& \frac{L}{\sqrt{2\pi(\mu+\lambda) T}}\int d\chi   e^{-S_\chi}
   \\
   &\times& \exp\left\{ \frac{L^2}{T} \left[-\frac{\chi^2 }{2(\mu+\lambda)} + i \chi \left(\xi^2-1\right)\right] \right\},
\nonumber
\EEA
where
 \BEA
 \nonumber
 S_\chi&=& \int \limits _{\mathbf q\omega}
 \left[ \frac{\rho\omega^2\! +\! \varkappa q^4-i\chi}{2} (|\mathbf h^{\mathbf q,\omega}|^2\!
+\! |\mathbf u^{\mathbf q,\omega}|^2 \!) \right.
 \\
 \label{Schi}
 &+& \left.
\mu | u_{\alpha \beta}^{\mathbf q,\omega}|^2 \!+\!
 \frac{\lambda}{2} | u_{\alpha \alpha}^{\mathbf q,\omega}|^2 \right]\! .
  \EEA

We will first discuss what happens in the harmonic approximation and later include the anharmonic coupling between the in-plane and out-of-plane modes. In the harmonic approximation,  $S_\chi$ simplifies,
\BEA
\nonumber
 S_\chi&=&
  \int \limits _{\mathbf q\omega}
 \left[ \frac{\rho\omega^2\! +\! \varkappa q^4-i\chi}{2}
 (|\mathbf h^{\mathbf q,\omega}|^2\! +    \! |\mathbf u^{\mathbf q,\omega}|^2 \!) \right.
 \\  \label{Schi1}
 &+&
\left.\frac{\mu\xi^2}{2} q^2  |\mathbf u^{\mathbf q,\omega}|^2+ \frac{(\mu+\lambda)\xi^2}{2}  |\mathbf q\mathbf u^{\mathbf q,\omega}|^2 \right].
  \EEA
After having performed here  the Gaussian integration over   $u$ and $h$, we are left with an integral over $d\chi$, which can be calculated by the stationary-phase method.  Denoting the value of $\chi$ obeying the stationary-phase condition as
$$
\chi_0=i\sigma_0,
$$
we express the free energy $F$ in terms of $\sigma_0$ and $\xi$:
\BEA
\frac{F}{L^2} &=& -\frac{\sigma_0^2}{2(\mu+\lambda)} +\sigma_0 (\xi^2-1)
\label{Fsigma}
\\ \nonumber
&+&
   \frac 12 \int \limits _{\mathbf q\omega}
  \left  \{\!d_c\ln \left( \varkappa q^4 +\sigma_0 q^2 + \rho \omega^2\right)\right.
\\\nonumber
&+& \ln\left[ \varkappa q^4 +(\sigma_0+\mu\xi^2)  q^2 + \rho \omega^2\right]
\\\nonumber
&+&\left. \ln \left[ \varkappa q^4 +(\sigma_0+[2\mu+\lambda]\xi^2 )q^2 + \rho \omega^2\right] \right\}.
\EEA

Effects of  the anharmonic coupling  between in-plane and flexural modes lead to the following modifications of Eq.~\eqref{Fsigma}.  First, the bending rigidity $\varkappa$ gets renormalized,  $\varkappa \to \varkappa_q$, as discussed in Sec.~\ref{sec.I}.  Second, there arises a self-energy correction \cite{Katsnelson2} $\sigma_1$  to $\sigma_0$ in the arguments of logarithms. We will calculate  $\sigma_1$ in Appendix ~\ref{anh}.  As discussed in Sec.~\ref{sec2a} and in Appendix~\ref{anh}, the total coefficient of the $q^2$ term in the phonon propagator is exactly the external tension $\sigma$,
\be
\label{sigma-total}
\sigma_0 +\sigma_1=\sigma.
\ee
This relation is a manifestation of a Ward identity that is verified within the RG analysis (in one-loop order) in Appendix~\ref{app-D}.

The stationary-point condition  $\p F/\p \sigma_0 =0$  yields
\BEA
&& \nonumber
\frac{\sigma_0}{ \mu+\lambda}\!=\!\xi^2-1\! +\!
 \frac 12 \int \limits _{\mathbf q\omega}
\left[ \frac{d_c q^2}{\varkappa q^4 \!+\!(\sigma_0\!+\!\sigma_1) q^2\! +\! \rho \omega^2}   \right.
\\   &&\left.  + \frac{q^2}{\!(\sigma_0+\sigma_1+\mu\xi^2)  q^2
+ \rho \omega^2}\!\right.  \nonumber
\\
&&\left.
+ \frac{q^2}{[\sigma_0
+ \sigma_1+(2 \mu+\lambda)\xi^2]  q^2\! +\! \rho \omega^2} \right].
 \label{stat-phase}
\EEA
Here, we have neglected the term $\varkappa q^4$ which is small compared to   $\mu\xi^2  q^2$ and  $(2\mu+\lambda) \xi^2  q^2.$   The stretching parameter $\xi$ entering this equation is fixed by the external tension $\sigma$,  which is given by
 a derivative of  the free energy with respect to the``projected area''  of the  sample,  $A=\xi^2L^2$ (see Eq.~\eqref{sigma-def-1}).
 Substituting   Eq.~\eqref{Fsigma} into Eq.~\eqref{sigma-def-1}, we get
 \BEA \nonumber
&&\sigma\!=\!\sigma_0
+
 \frac 12 \int \limits _{\mathbf q\omega}
\left[ \frac{\mu q^2}{\!(\sigma_0+\sigma_1+\mu\xi^2)  q^2 \!+\! \rho \omega^2} \right.
\\&&\left.  +\frac{(2\mu+\lambda)q^2}{[\sigma_0\!+\!\sigma_1+(2 \mu+\lambda)\xi^2]  q^2\! +\! \rho \omega^2} \right].
 \label{sigma-F}
\EEA
Neglecting in the denominator the tension $\sigma_0 +\sigma_1=\sigma$ (which is assumed to be much smaller than the in-plane elastic moduli $\mu$ and $\lambda$), we get
\be
\sigma=\sigma_0+\delta \sigma,
\label{sigma0+sigma1}
\ee
where
 \be
 \label{sigma-F1}
 \delta \sigma=
 \frac 12 \int \limits _{\mathbf q\omega}
\left[ \frac{\mu q^2}{\!\mu\xi^2  q^2 \!+\! \rho \omega^2} +\frac{(2\mu+\lambda)q^2}{(2 \mu+\lambda)\xi^2  q^2\! +\! \rho \omega^2} \right].
\ee
We will show in Appendix~\ref{anh} that the one-loop contributions $\sigma_1$ and  $\delta \sigma$ arising in the analysis of the tension on the basis of the FP Green function and of the thermodynamic relation,   Eq.~\eqref{sigma-def-1}, respectively, are identical, $\sigma_1 = \delta\sigma$, so that Eqs.~(\ref{sigma-total}) and (\ref{sigma0+sigma1}) are fully consistent.

\section{Anharmonic coupling between in-plane and out-of-plane phonons} \label{anh}

In this Appendix, we provide a derivation of the self-energy correction $\sigma_1$ in Eq.~(\ref{stat-phase}), which is generated by the
anharmonic coupling between in-plane and out-of-plane modes.  In combination with results of Appendix~\ref{F-S}, this allows us to derive equations (\ref{Fsigma0}) and (\ref{xi})  the main text used there for the analysis of thermomechanical properties of the membrane.

We begin with the action (\ref{Schi}). Integrating out the in-plane modes, we get an energy functional which only depends on $h-$fields:  \cite{Katsnelson2}
\begin{widetext}
\be
\label{Eh}
E[h]=\frac 12\int \limits _{\mathbf q\Omega}(\varkappa q^4 +\rho\omega^2)|\mathbf h^{\mathbf q\omega} |^2+\frac 18\int \limits _{\mathbf q\Omega} \int \limits _{\mathbf Q \omega}\int \limits _{\mathbf Q' \omega'}
R^{\alpha\beta\gamma\theta}(\mathbf q, \Omega) (Q_\alpha-q_\alpha) Q_\beta  (Q'_\gamma +q_\gamma) Q' _\theta
 (\mathbf h_\mathbf Q^\omega
\mathbf  h_{\mathbf q-\mathbf Q}^{ \Omega-\omega}) (\mathbf h_{\mathbf Q'}^{\omega'}
\mathbf  h_{-\mathbf q-\mathbf Q'}^{ -\Omega-\omega'})
\ee
   Here $R^{\alpha\beta\gamma\theta}(\mathbf q, \Omega)$ is  the coupling tensor    with the following  nonzero components [in the basis of vectors $\mathbf n=(-q_y, q_x)/|q|, $ $\mathbf m= \mathbf q/|q|$]:
\BEA
&& R^{nnnn}= \frac{4\mu(\mu+\lambda)}{2 \mu+\lambda}
+\frac{\rho \Omega^2 \lambda^2}{(2\mu+\lambda)[q^2(2\mu+\lambda) \xi^2 +\rho\Omega^2] }, \\
&& R^{mmmm}= \frac{\rho \Omega^2 (2\mu +\lambda)}{q^2(2\mu+\lambda)\xi^2 +\rho\Omega^2 }, \\
&& R^{mmnn}=R^{nnmm}= \frac{\rho \Omega^2 \lambda}{q^2(2\mu+\lambda)\xi^2 +\rho\Omega^2 },\\
&& R^{mnmn}= \frac{4\rho \Omega^2 \mu}{q^2\mu\xi^2 +\rho\Omega^2 }.
\EEA
\end{widetext}
The $R^{\alpha \beta\gamma\theta}$ couplings characterize the quartic interaction of the out-of-plane modes. This interaction generates self-energies the $hh$  correlation functions.
The coupling constant  $R^{nnnn}$  leads to  a self-energy that scales as $q^4\ln q$ and, therefore,     is responsible for the power-law renormalization of $\varkappa.$  On the other hand, the
couplings $R^{mmmm}$ and $R^{mnmn}$ lead to self-energy corrections that scale as $q^2$.
%
%
Specifically, the self-energy originating from  the coupling $R^{mmmm}$ reads
\BEA \nonumber
&&\Sigma^{mmmm}_{\mathbf Q,\Omega}=\int\limits_{\mathbf q \omega}\frac{(\mathbf Q\mathbf q-q^2)^2 (\mathbf Q\mathbf q)^2}{q^4}  \frac{\rho (2\mu +\lambda)\omega^2}{(2\mu+\lambda) q^2\xi^2 +\rho \omega^2}
\\
&&\times \frac{1}{\varkappa (\mathbf q- \mathbf Q)^4 +\rho (\omega-\Omega)^2}
\EEA
(the factor $1/8$ in the coupling is cancelled due to $8$ pairing possibilities of $h$ fields.)
In the limit $\Omega \to 0$ $Q \to 0,$ we get
\be
\Sigma^{mmmm}_{Q\to 0,\Omega\to0}=\frac{Q^2}{2}\int\limits_{\mathbf q \omega}  \frac{\rho (2\mu +\lambda) q^2\omega^2}{(2\mu+\lambda) \xi^2q^2 +\rho \omega^2}\frac{1}{\varkappa  q^4 +\rho \omega^2}
\ee
The integral is determined by the ultraviolet cut-off of the theory $q_{\rm uv}$.
 For $q< q_{\rm uv},$
one can neglect  the term $\varkappa  q^4$ in the denominator.
 This yields
 $$\Sigma^{mmmm} \simeq Q^2\sigma_1^{mmmm},$$
 with
\be
\sigma_1^{mmmm} \approx \frac{1}{2}\int\limits_{\mathbf q \omega}  \frac{ (2\mu +\lambda) q^2}{(2\mu+\lambda)\xi^2 q^2 +\rho \omega^2}.
\ee
Proceeding in the same way, we find
\be
\sigma_1^{mnmn} \approx \frac{1}{2}\int\limits_{\mathbf q \omega}  \frac{ \mu  q^2}{\mu\xi^2 q^2 +\rho \omega^2}.
\ee
Combining these contributions, we get the following result for the total one-loop coefficient
$$\sigma_1=\sigma_1^{mmmm}+\sigma_1^{mnmn}$$ of the $Q^2$ self-energy:
\be
\sigma_1 \approx \frac{1}{2}\int\limits_{\mathbf q \omega}q^2\left[  \frac{ 2\mu +\lambda}{(2\mu+\lambda)\xi^2 q^2 +\rho \omega^2}
+\frac{ \mu  }{\mu \xi^2 q^2 +\rho \omega^2}.
\right]
\ee
Comparing this equation with Eq.~\eqref{sigma-F1}, we satisfy ourselves that
\be
\sigma_1=\delta \sigma,
\label{s1ds}
\ee
as was stated in the end of Appendix~\ref{F-S}. In combination with Eq.~(\ref{sigma0+sigma1}), this yields Eq.~(\ref{sigma-total}).

 We have thus explicitly demonstrated that Eq.~\eqref{stat-phase} can be written
 in terms of the applied tension $\sigma$,
\be  \label{hooke}
\frac{\sigma}{ \mu+\lambda}\!=\!\xi^2-1\! +\!
 \frac 12 \int \limits _{\mathbf q\omega}
 \frac{d_c q^2}{\varkappa q^4 \!+\!\sigma q^2\! +\! \rho \omega^2}   +a,
\ee
where
\BEA
 &&a=  \frac {1}{2(\mu+\lambda)} \int \limits _{\mathbf q\omega}q^2 \left[\frac{2\mu+\lambda}{\!(\sigma+\mu\xi^2)  q^2 \!+\! \rho \omega^2}\! \right.
 \\ \nonumber
 && \left. +\!\frac{3\mu+2 \lambda}{[\sigma+(2 \mu+\lambda)\xi^2]  q^2\! +\! \rho \omega^2}\right]
 \label{a}
\EEA
is an ultraviolet correction that can be fully absorbed in the renormalization of $\xi.$  Equation \eqref{hooke} is the generalized
Hooke's law.  We emphasize once more that the denominator in the r.h.s. of   Eq.~ \eqref{hooke} contains only the external tension  $\sigma$ and is not sensitive to the ultraviolet cutoff of the theory.

    Since $a$ depends on $T,$  it leads to  a correction to $\alpha_T.$ One can show that  this correction is of the order of
    $$
    \delta \alpha_T \sim \frac{1}{\varkappa}  \left(\frac{T}{g \varkappa}\right)^2,
    $$
     and is thus small  in comparison with Eq.~\eqref{alphaT-q} under the condition  $T<g\varkappa\sqrt{2/\eta +\ln(1/g)}$ (for graphene $T< 1000$ K).
     Therefore, one can safely discard this contribution for not too high temperatures, $T < T_{\rm uv}$, with $T_{\rm uv}$ given by
     Eq.~(\ref{condition-T}), which is the temperature range of our interest in this paper. Neglecting $a$ in Eq.~(\ref{hooke}), we get Eq.~\eqref{xi} of the main text.

Using the identity (\ref{sigma-total}), we can also cast the free energy of renormalized out-of-plane modes in Eq.~(\ref{Fsigma})
in the form
     \BEA
\frac{F}{L^2}&=& -\frac{\sigma^2}{2B_0} +\sigma (\xi^2-1)
 \label{Fsigma0-app}
\\ \nonumber
&+&
   \frac {d_{\rm c}}{2} \sum \limits _{\mathbf q\omega}
   \ln \left( \varkappa_q q^4 +\sigma q^2 + \rho \omega^2\right),
\EEA
which is Eq.~(\ref{Fsigma0}) of the main text.

  \section{Quantum renormalization group. \label{ren}}

      In this Appendix, we derive the quantum RG equations that are presented in Sec.~\ref{sec-RG} of the main text. An alternative derivation is presented in Appendix~\ref{app-D}.

The quantum renormalization grouo operates in the region of momenta $q_T < q < q_{\rm uv}$.  For momenta $q$ below $q_{\rm uv} \sim \sqrt{\mu/\varkappa}$, the flexural phonons are  softer than the in-plane modes: $\omega_q < \omega_q ^{\perp,  \parallel}$.  Here $\omega_q ^{\perp} =\sqrt{\mu/\rho}q$ and
$\omega_q ^{\parallel}=\sqrt{(2\mu+\lambda)/\rho}q$.  Since in the considered region of momenta $\hbar  \omega_q  \gg T $, the flexural phonons are frozen out, so that the relevant RG equations are of zero-temperature character \cite{kats2}.

Here,  we derive RG equations  by  using the energy functional Eq.~\eqref{Eh} where in-plane modes have been integrated out.

We have  demonstrated above that  the terms  scaling as $q^2$ in the flexural phonons self-energy  cancel (for zero external tension $\sigma=0$).   After this cancellation is taken into account, all  remaining   effects  related to retardation turn out to be small  for $ q< q_{\rm uv}$  and can be safely  neglected.  Hence, it is sufficient to keep  the only component of the interaction tensor $R^{\alpha\beta\gamma\theta},$
\be
R^{nnnn} \approx\frac{4\mu(\mu+\lambda)}{2 \mu+\lambda}=Y.
\ee
 To proceed, we use  the  approach analogous to one  developed in   Ref.~\onlinecite{my-crump} for high-temperature case.    To find the renormalization  of elastic coefficients within this approach, we have to calculate the polarization operator and  the self-energy of $h-$fields.
 The bare Green function for $h$-field reads
 \be
 G^0_{\omega, \mathbf k}=\frac{1}{\varkappa k^4 +\rho \omega^2}.
  \ee
The polarization operator is given by  the following equation
\be
\Pi_{\Omega, \mathbf q}=\frac{d_{\rm c}}{3}\int \frac{d \omega d^2\mathbf k}{(2\pi)^3} k_\perp^4 G^0_{\omega, \mathbf k },
G^0_{\Omega- \omega, \mathbf q-\mathbf k },
\label{polariz}
\ee
where $\mathbf k_\perp=[\mathbf k \times \mathbf q]/q$. Equation  (\ref{polariz})
is a quantum counterpart of Eq.~(37) of Ref.~\onlinecite{my-crump} derived there for the high-temperature classical regime.  Performing the integration over $d\mathbf \omega$, we get
\be
\Pi_{\Omega, \mathbf q}=\frac{d_{\rm c}}{6\rho^2}\int \frac{ d^2\mathbf k}{(2\pi)^2}   \frac{k_\perp^4(\omega^0_{\mathbf k-\mathbf q}+\omega^0_{\mathbf k})}{ \omega^0_{\mathbf k-\mathbf q}\omega^0_{\mathbf k}  [(\omega^0_{\mathbf k-\mathbf q}+\omega^0_{\mathbf k})^2 +\Omega^2]}
\ee
Carrying out the remaining momentum integration we find, in the limit $\Omega = 0$ and $q\to 0$,
\be
\Pi_\mathbf q =\Pi_{\Omega = 0 , \mathbf q \to 0}\approx \frac{d_{\rm c}}{64 \pi \rho^{1/2} \varkappa^{3/2}} \ln \left(\frac{q_{\rm uv}}{q}\right).
\label{Pi}
\ee
Next, we use Eqs. (33) and (34) of Ref.~\onlinecite{my-crump} to find screening of coupling constants $Y $ and $\mu:$
\BEA
&&Y_\mathbf q = \frac{Y}{1+3Y \Pi_\mathbf q/2}\approx Y-3Y^2 \Pi_\mathbf q/2 ,
\label{N}
\\
&&\mu_\mathbf q = \frac{\mu}{1+2\mu \Pi_\mathbf q}\approx \mu-2\mu^2 \Pi_\mathbf q.
\label{mu}
\EEA
We notice that for $D=2$  the interaction between $h-$fields in Eq.~\eqref{Eh}  depends on coupling  $Y=R^{nnnn} (\Omega\to 0) $  only, while $\mu$ drops out from  Eq.~\eqref{Eh}.   Therefore,  in order to obtain  Eq.~\eqref{mu},  one should first consider $D\neq 2,$ and then take the limit $D\to 2.$

Substituting  Eq.~\eqref{Pi} into Eqs.~\eqref{N} and \eqref{mu}, we find RG equations for  in-plane elastic moduli:
\BEA
\frac{dY}{d\Lambda}&=&-\frac{3 d_{\rm c}Y^2}{128\pi\rho^{1/2} \varkappa^{3/2}}, \\
\frac{d(\mu +\lambda) }{d\Lambda}&=&-\frac{d_{\rm c}(\mu+\lambda)^2}{16\pi\rho^{1/2} \varkappa^{3/2}},
\EEA
where  $ \Lambda=\ln\left( {q_{\rm uv}}/{q}\right).$ These equations are equivalent to Eqs.~(9) and (10) of Ref.~\onlinecite{kats2}.
Renormalization of self-energy of $h-$field is given by an equation similar to Eq.~(43) of Ref.~\onlinecite{my-crump}:
\be
\Sigma_{\omega,\mathbf k}= \int \frac{d \Omega d^2\mathbf q}{(2\pi)^3} k_\perp^4 Y_\mathbf q G^0_{\omega-\Omega, \mathbf k-\mathbf q }.
\ee
Integrating over $d\Omega,$  taking the limit $\omega \to 0,$ and neglecting the dependence of $Y$ on $q,$ we get
\be
\Sigma_{\omega \to 0,\mathbf k}= \frac{Y}{2 \sqrt{\varkappa \rho}}\int \frac{ d^2\mathbf q}{(2\pi)^2} \frac{k_\perp^4}{|\mathbf k-\mathbf q|^2} .
\label{eq:sigmaV:i}
\ee
A straightforward analysis of this integral shows that $\Sigma$ scales as $k^4\ln k,$ which implies a renormalization of $\varkappa$,
\be
\frac{d\varkappa}{d \Lambda}= \frac{3Y}{32\pi \sqrt{\varkappa \rho}}.
\label{kappa}
\ee
 The Eq.~\eqref{kappa}  coincides up to the sign  with Eq.~(11) of Ref.~\onlinecite{kats2}. From Eqs.~\eqref{N},\eqref{mu} and  \eqref{kappa}, one easily obtains  Eqs.~ \eqref{RG-Y} and \eqref{kappa0} of the main text, with $g$ given by Eq.~\eqref{g0}.

\begin{widetext}

 \section{Background-field renormalization} \label{app-D}

In this Appendix, we perform a derivation of quantum RG equations (Sec.~\ref{sec-RG} of the main text) alternative to that presented in Appendix~\ref{ren}.
For this purpose, we evaluate the self-energies of the propagators of in- and out-of-plane phonons
within one-loop approximation by using the approach of Ref.~[\onlinecite{kats2}].

We start from the Lagrangian given by Eq. \eqref{Luh}. Using definition  \eqref{sigma-def-1}, we obtain formally exact relation between the external tension and the global stretching factor:
\begin{gather}
\frac{\sigma}{B} =\xi^2-1 +  \int\limits_{\bm{q},\omega} q^2 \Biggl ( \frac{d_{\rm c}}{2} G_{\omega,q} +  \frac{2\mu+\lambda}{2B}  F^{(t)}_{\omega,q}
+ \frac{3\mu+2\lambda}{2B} F^{(l)}_{\omega,q} \Biggr )
 + \frac{1}{2\xi B}(\mu\delta_{\gamma\beta}\delta_{\eta\alpha}+ \lambda \delta_{\gamma\alpha}\delta_{\eta\beta}) \left \langle \partial_\alpha u_\gamma \bigl ( \partial_\eta \bm{u} \partial_\beta \bm{u} + \partial_\eta \bm{h} \partial_\beta \bm{h}\bigr )\right \rangle .
\label{eq:sf3:i}
\end{gather}
Here $F^{(t,l)}_{\omega,q}$ and  $G_{\omega,q}$ are exact (with  respect to the full Lagrangian \eqref{Luh}) propagators of in- and out-of plane phonons, respectively:
\begin{gather}
\langle u_\alpha(\bm{q},i\omega) u_\beta(-\bm{q},-i\omega)\rangle = F^{(l)}_{\omega,q} \frac{q_\alpha q_\beta}{q^2} + F^{(t)}_{\omega,q} \left (\delta_{\alpha\beta} - \frac{q_\alpha q_\beta}{q^2} \right ), \qquad
\langle \bm{h}(\bm{q},i\omega) \bm{h}(-\bm{q},-i\omega)\rangle = d_{\rm c} G_{\omega,q} .
\end{gather}
The exact propagators can be cast in the following form
\begin{equation}
\begin{split}
[F^{(l)}_{\omega,q}]^{-1}  &  = \rho \omega^2 +[(2\mu+\lambda)\xi^2 + (\mu+\lambda)(\xi^2-1)]q^2+\varkappa q^4  - \Sigma^{(l)}_{\omega,q}\  , \\
[F^{(t)}_{\omega,q}]^{-1} &  = {\rho \omega^2 +[\mu\xi^2+(\mu+\lambda)(\xi^2-1)] q^2+\varkappa q^4 -\Sigma^{(t)}_{\omega,q}} \ , \\
[G_{\omega,q}]^{-1} & = {\rho \omega^2 + (\mu+\lambda)(\xi^2-1) q^2+\varkappa q^4 - \Sigma_{\omega,q}} \ ,
\end{split}
\label{eq:prop:i}
\end{equation}
where the self-energies $\Sigma^{(l,t)}_{\omega,q}$ and $\Sigma_{\omega,q}$ take into account interaction of in- and out-of-plane modes encoded in the Lagrangian \eqref{L0}. The expressions \eqref{eq:prop:i} in the absence of self-energies corresponds to the Gaussian part of the Lagrangian \eqref{Luh}.
We emphasize the appearance of linear in $q^2$ term in the propagator of the out-of-plane phonon due to the linear in $K$ term in the Lagrangian \eqref{Luh}. As we will see below it will be compensated by the linear in $q^2$ term from  the self-energy $\Sigma_{\omega,q}$.
 To avoid confusion, we note that the definition of the self-energies used in  Appendixes  \ref{F-S}, \ref{anh}, and \ref{ren} is different   compared to the definition which we use here.    Of course, this does not change  the physical propagators and, in particular, the cancelation of  $\propto q^2$ contributions to the inverse propagator of out-of-plane phonons in the absence of the external stress ($\sigma=0$). This statement   can be written  as $\sigma_0+\sigma_1=0$ (as was done in Appendixes  \ref{F-S}, \ref{anh},  \ref{ren}), or, equivalently,  as  $ B(\xi^2-1) - \lim\limits_{q\to 0}\Sigma_{\omega=0,q} / q^2=0$ within background-filed  renormalization approach used in this Appendix.
 %
%

In order to find the corresponding self-energies $\Sigma^{(l,t)}_{\omega,q}$ and $\Sigma_{\omega,q}$, we use the background field method. We split the fields $\bm{u}$ and $\bm{h}$ on slow $\bm{u^\prime}$, $\bm{h^\prime}$ and fast $\bm{\tilde{u}}$, $\bm{\tilde{h}}$ components in the momentum and frequency spaces, $\bm{u} =\bm{u^\prime}+ \bm{\tilde{u}}$ and $\bm{h}=\bm{h^\prime}+\bm{\tilde{h}}$. We denote the corresponding momentum scale which separates fast and slow modes as $q^{\rm sf}_{\Lambda}$. Then the interaction terms in the Lagrangian \eqref{L0} generates the following interaction terms between slow and fast components. For a sake of simplicity, we consider the case of $d_{\rm c}=1$ and restore arbitrary dimensionality in the final results for the self-energies only. Then, limiting ourselves to the first and second orders in slow components, we find
\begin{align}
S_{\bm{u^\prime},\tilde{h}}^{(1),2} & = \int_0^\beta d\tau \int d^2 \bm{x} \Bigl ( \mu \xi \partial_\alpha u^\prime_\beta + \frac{\lambda \xi}{2} \delta_{\alpha\beta} \partial_\eta u^\prime_\eta \Bigr ) \partial_\alpha {\tilde{h}} \partial_\beta {\tilde{h}} , \notag \\
S_{{h^\prime},\bm{\tilde{u}},\tilde{h}}^{(1),1,1}& = \int_0^\beta d\tau \int d^2 \bm{x} \Bigl [ \mu \xi \Bigl (\partial_\alpha \tilde{u}_\beta + \partial_\beta \tilde{u}_\alpha\Bigr ) + \lambda \xi \delta_{\alpha\beta} \partial_\eta \tilde{u}_\eta \Bigr ] \partial_\alpha {h^\prime} \partial_\beta {\tilde{h}}, \notag \\
S_{\bm{u^\prime},\bm{\tilde{u}}}^{(1),2} & = \int_0^\beta d\tau \int d^2 \bm{x} \Bigl [\Bigl ( \mu \xi \partial_\alpha u^\prime_\beta + \frac{\lambda \xi}{2} \delta_{\alpha\beta} \partial_\eta u^\prime_\eta \Bigr ) \partial_\alpha \tilde{u}_\gamma \partial_\beta \tilde{u}_\gamma +  \Bigl ( \mu \xi \Bigl (\partial_\alpha \tilde{u}_\beta + \partial_\beta \tilde{u}_\alpha\Bigr ) + \lambda \xi \delta_{\alpha\beta} \partial_\eta \tilde{u}_\eta \Bigr ) \Bigr ] \partial_\alpha u^\prime_\gamma \partial_\beta \tilde{u}_\gamma, \notag \\
S_{\bm{u^\prime},\bm{\tilde{u}}}^{(2),2} & =\frac{1}{2} \Bigl ( \mu \delta_{\alpha\theta}\delta_{\beta \eta} + \frac{\lambda}{2} \delta_{\alpha\eta}\delta_{\beta \theta}\Bigr ) \int_0^\beta d\tau \int d^2 \bm{x} \Bigl (
\partial_\alpha u^\prime_\gamma \partial_\eta u^\prime_\gamma \partial_\theta \tilde{u}_\zeta \partial_\beta \tilde{u}_\zeta + \partial_\alpha u^\prime_\gamma \partial_\eta \tilde{u}_\gamma \partial_\theta \tilde{u}_\zeta \partial_\beta {u}^\prime_\zeta+ \partial_\alpha u^\prime_\gamma \partial_\eta \tilde{u}_\gamma \partial_\theta {u}^\prime_\zeta \partial_\beta \tilde{u}_\zeta\Bigr ) , \notag \\
S_{{h^\prime},{\tilde{h}}}^{(2),2} & =\frac{1}{4} \Bigl ( \mu \delta_{\alpha\theta}\delta_{\beta \eta} + \frac{\lambda}{2} \delta_{\alpha\eta}\delta_{\beta \theta}\Bigr ) \int_0^\beta d\tau \int d^2 \bm{x} \Bigl (
2 \partial_\alpha h^\prime \partial_\eta h^\prime \partial_\theta \tilde{h} \partial_\beta \tilde{h}
+
 \bigl (\partial_\alpha h^\prime \partial_\eta \tilde{h}+ \partial_\alpha  \tilde{h} \partial_\eta h^\prime \bigr )
\bigl (  \partial_\theta \tilde{h} \partial_\beta {h}^\prime+\partial_\theta {h}^\prime \partial_\beta \tilde{h}\bigr ) \Bigr ), \notag \\
S_{\bm{u^\prime},{\tilde{h}}}^{(2),2} & =\frac{1}{2} \Bigl ( \mu \delta_{\alpha\theta}\delta_{\beta \eta} + \frac{\lambda}{2} \delta_{\alpha\eta}\delta_{\beta \theta}\Bigr ) \int_0^\beta d\tau \int d^2 \bm{x}\,
\partial_\alpha u^\prime_\gamma \partial_\eta u^\prime_\gamma \partial_\theta \tilde{h} \partial_\beta \tilde{h}, \notag \\
S_{{h^\prime},\bm{\tilde{u}}}^{(2),2} & =\frac{1}{2} \Bigl ( \mu \delta_{\alpha\theta}\delta_{\beta \eta} + \frac{\lambda}{2} \delta_{\alpha\eta}\delta_{\beta \theta}\Bigr ) \int_0^\beta d\tau \int d^2 \bm{x}\,
\partial_\alpha h^\prime \partial_\eta h^\prime \partial_\theta {\tilde{u}}_\gamma  \partial_\beta {\tilde{u}}_\gamma, \notag \\
S_{h^\prime,\bm{u^\prime},\tilde{h},\bm{\tilde{u}}}^{(1,1),1,1}  & =\frac{1}{2} \Bigl ( \mu \delta_{\alpha\theta}\delta_{\beta \eta} + \frac{\lambda}{2} \delta_{\alpha\eta}\delta_{\beta \theta}\Bigr ) \int_0^\beta d\tau \int d^2 \bm{x}\, \bigl (\partial_\alpha {u^\prime}_\gamma \partial_\eta {\tilde{u}}_\gamma+ \partial_\alpha  {\tilde{u}}_\gamma \partial_\eta  {u^\prime}_\gamma \bigr ) (  \partial_\theta \tilde{h} \partial_\beta {h}^\prime+\partial_\theta {h}^\prime \partial_\beta \tilde{h}\bigr ) .
 \end{align}
After integration over fast variables we find the correction to the Gaussian part of the action for the in-plane ($\delta S_u^{(2)}$) and out-of-plane ($\delta S_h^{(2)}$) slow modes:
\begin{gather}
\delta S_u^{(2)}  = \langle  S_{\bm{u^\prime},{\tilde{h}}}^{(2),2} \rangle_0+ \langle S_{\bm{u^\prime},\bm{\tilde{u}}}^{(2),2} \rangle_0  - \frac{1}{2} \left \langle \Bigl [S_{\bm{u^\prime},\bm{\tilde{u}}}^{(1),2}  \Bigr ]^2 \right \rangle_0 - \frac{1}{2} \left \langle \Bigl [ S_{\bm{u^\prime},\tilde{h}}^{(1),2}\Bigr ]^2 \right \rangle_0 , \label{eq:dSu:i}\\
\delta S_h^{(2)} = \langle  S_{{h^\prime},{\tilde{h}}}^{(2),2} \rangle_0 + \langle S_{{h^\prime},\bm{\tilde{u}}}^{(2),2} \rangle_0 - \frac{1}{2} \left \langle \Bigl [S_{{h^\prime},\bm{\tilde{u}},\tilde{h}}^{(1),1,1}  \Bigr ]^2 \right \rangle_0  .
\label{eq:dSh:i}
\end{gather}
Here the average $\langle \dots \rangle_0$ is with respect to the Gaussian part of the full Lagrangian \eqref{Luh}. The self-energies can be found from the following expressions
\begin{gather}
\delta S_u^{(2)} = - \frac{1}{2} \int\limits_{\bm{q},\omega} u^\prime_\alpha(\bm{q},i\omega) u^\prime_\beta(-\bm{q},-i\omega) \left [ \Sigma^{(l)}_{\omega,q} \frac{q_\alpha q_\beta}{q^2} + \Sigma^{(t)}_{\omega,q} \left (\delta_{\alpha\beta} - \frac{q_\alpha q_\beta}{q^2} \right ) \right ],
\notag \\
\delta S_h^{(2)} = - \frac{1}{2} \int\limits_{\bm{q},\omega}  \bm{h^\prime}(\bm{q},i\omega) \bm{h^\prime}(-\bm{k},-i\omega) \Sigma_{\omega,q} .
\label{eq:Su:i}
\end{gather}

Evaluation of averages in $\delta S_h^{(2)}$ yields
\begin{gather}
\Sigma_{\omega,q} = - q^2 \int\limits_{\bm{k},\Omega} \Biggl [(2\mu+\lambda) k^2 G^{(0)}_{\Omega,k}+ \frac{\lambda+\mu}{2} k^2 \Bigl (F^{(l),(0)}_{\Omega,k} + F^{(t),(0)}_{\Omega,k}\Bigl )\Biggl ] -
\int\limits_{\bm{k},\Omega} \xi^2 G^{(0)}_{\omega+\Omega,\bm{q}+\bm{k}} \Bigl \{ \Bigl [
(2\mu+\lambda)^2 k^2 (\bm{q}\cdot \bm{k})^2 \notag \\
+4\mu(2\mu+\lambda) (\bm{q}\cdot \bm{k})^3+2\lambda(2\mu+\lambda) (\bm{q}\cdot \bm{k}) q^2 k^2+ 4\mu^2 (\bm{q}\cdot \bm{k})^4 k^{-2} + 4\mu\lambda (\bm{q}\cdot \bm{k})^2 q^2+\lambda^2 k^2q^4
\Bigr ] F^{(l),(0)}_{\Omega,k} \notag \\
+ \mu^2 \frac{[\bm{k}\bm{\times}\bm{q}]^2}{k^2}\bigl (k^2 + 2 (\bm{k}\cdot \bm{q})^2 \bigl )  F^{(t),(0)}_{\Omega,k}\Bigr \} .
\label{eq:C:i}
 \end{gather}
Here $F^{(l,t),(0)}_{\Omega,k}$ and $G^{(0)}_{\Omega,k}$ denote the propagators of in- and out-of-plane modes within the Gaussian approximation to the full Lagrangian \eqref{Luh}. In Eq. \eqref{eq:C:i} the terms linear in the propagators corresponds to the contributions from $\langle  S_{{h^\prime},{\tilde{h}}}^{(2),2} \rangle_0$ and $\langle S_{{h^\prime},\bm{\tilde{u}}}^{(2),2} \rangle_0$ whereas the terms proportional to the product of propagators for the in-plane and out-of-plane modes corresponds to the last contribution
in the right hand side of Eq. \eqref{eq:dSh:i}. We note that the first and last terms in the right hand side of Eq. \eqref{eq:C:i} corresponds to the self-energy contribution due to effective interaction tensor $R^{\alpha\beta\gamma\theta}$. The second term in the right hand side of Eq. \eqref{eq:C:i} appears due to the interaction of two flexural phonons with two in-plane phonons. This vertex is not included in the interaction tensor $R^{\alpha\beta\gamma\theta}$. However, as we shall demonstrate below, this interaction is taken into account in the approach of Appendix \ref{F-S}.

In the limit $q\to 0$ and $\omega=0$ we find from Eq. \eqref{eq:C:i}
\begin{equation}
\Sigma_{\omega=0,q}= - q^2 \int\limits_{\bm{k},\omega} \Biggl [(2\mu+\lambda) k^2 G^{(0)}_{\Omega,k}+ \frac{\lambda+\mu}{2} k^2 \Bigl (F^{(l),(0)}_{\Omega,k} + F^{(t),(0)}_{\Omega,k}\Bigl )\Biggl ]+
\frac{q^2\xi^2}{2}  \int\limits_{\bm{k},\Omega} k^4 G^{(0)}_{\Omega,k}\Bigl [ (2\mu+\lambda)^2 F^{(l),(0)}_{\Omega,k}+ \mu^2 F^{(t),(0)}_{\Omega,k} \Bigr ] .
\label{eq:C01:i}
\end{equation}
It is convenient to regroup various terms in Eq. \eqref{eq:C01:i} in the following way:
\begin{equation}
\Sigma_{\omega=0,q}= - q^2 \frac{\mu+\lambda}{2} \int\limits_{\bm{k},\omega} k^2 \Biggl [G^{(0)}_{\Omega,k}+ F^{(l),(0)}_{\Omega,k} + F^{(t),(0)}_{\Omega,k}\Biggl ]+
\frac{q^2}{2}  \int\limits_{\bm{k},\Omega} k^2 G^{(0)}_{\Omega,k}\Bigl [ (2\mu+\lambda)^2\xi^2 k^2 F^{(l),(0)}_{\Omega,k}+ \mu^2 \xi^2 k^2 F^{(t),(0)}_{\Omega,k} - (3\mu+\lambda)\Bigr ] .
\label{eq:C012:i}
\end{equation}
As one can check, Eq. \eqref{eq:C012:i} can be written as  $\Sigma_{\omega=0,q}=[ \sigma_0+\sigma_1-B(\xi^2-1)] q^2 $. Using the precise form of the Gaussian propagators the result \eqref{eq:C012:i} can be equivalently rewritten as follows
\begin{equation}
\Sigma_{\omega=0,q}  = - \frac{q^2}{2} \int\limits_{\bm{k},\Omega} k^2 \Biggl \{ d_{\rm c} (\mu+\lambda) G^{(0)}_{\Omega,k}+(3\mu+2\lambda)F^{(l),(0)}_{\Omega,k}    +(2 \mu+\lambda) F^{(t),(0)}_{\Omega,k}  \Biggr \}  .
\label{eq:C1:i}
\end{equation}
Here we restore arbitrary value of $d_{\rm c}$. Comparing this result with the expression \eqref{eq:sf3:i} evaluated within the Gaussian theory, we conclude that within one-loop approximation the following identity holds
\begin{equation}
\sigma = B(\xi^2-1) - \lim\limits_{q\to 0}\Sigma_{\omega=0,q} / q^2 .
\label{eq:WI:i}
\end{equation}
Although at present we cannot prove this relation beyond the one-loop approximation, we believe that it should be satisfied in general (see discussion in the main text).

Expansion of the self-energy \eqref{eq:C:i} to the second in $q^2$ determines the one-loop renormalization of the bending rigidity:
\begin{equation}
\varkappa^\prime = \varkappa - \frac{1}{4!} \frac{\partial^4}{\partial q^4} \Sigma_{\omega=0,q} \Biggl |_{q=0} .
\end{equation}
As one can check by inspection of various terms in Eq. \eqref{eq:C:i}, $\partial^4 \Sigma_{\omega=0,q}/\partial q^4$, the logarithmically divergent contributions appear only for external momentum scale $q_T < q < q_{\rm uv} \sim \sqrt{\min\{\lambda,\mu\}/\varkappa}$.  Simplifying Eq. \eqref{eq:C:i} in this regime, we obtain Eq. \eqref{eq:sigmaV:i} where the integration over momentum is limited to $k>q^{\rm sf}_{\Lambda}$ whereas $q\ll q^{\rm sf}_{\Lambda}$. Performing integration over momentum, we find the following RG equation:
\begin{equation}
\frac{d \varkappa}{d\Lambda} = \frac{3 d_{\rm c}}{8 \pi\rho^{1/2}\varkappa^{1/2}}\frac{\mu(\mu+\lambda)}{2\mu+\lambda} \ ,
\label{eq:varkappa-ren-i}
\end{equation}
where $\Lambda= \ln q_{\rm uv}/q$ since the minimal value of $q^{\rm sf}_{\Lambda}$ is given by the external momentum $q$.  This equation coincides with Eq. \eqref{kappa0} in the main text.

\begin{table}
\caption{\label{Tab1}
The linear in $q^2$ contributions to $\Sigma^{(l,t)}_{\omega=0,q}$ from different terms in Eq. \protect\eqref{eq:dSu:i}.}
\begin{ruledtabular}
\begin{tabular}{c|cc}
& $-\Sigma^{(t)}_{\omega=0,q}/q^2$ & $-(\Sigma^{(l)}_{\omega=0,q}-\Sigma^{(t)}_{\omega=0,q})/q^2$  \\
\hline
$\langle  S_{\bm{u^\prime},{\tilde{h}}}^{(2),2} \rangle$  &  $\frac{\mu+\lambda}{2} \int_{\bm{k},\Omega} k^2 G^{(0)}_{\Omega,k}$ & 0 \\
$ \langle S_{\bm{u^\prime},\bm{\tilde{u}}}^{(2),2} \rangle$ & $\int_{\bm{k},\Omega} k^2 \Bigl ( \frac{9\mu+5\lambda}{8} F^{(l),(0)}_{\Omega,k}
 + \frac{11\mu+7\lambda}{8} F^{(t),(0)}_{\Omega,k}\Bigr )$ & $  \frac{\mu+\lambda}{4} \int_{\bm{k},\Omega} k^2  \Bigl (F^{(l),(0)}_{\Omega,k}-F^{(t),(0)}_{\Omega,k}\Bigr )$ \\
 $- \frac{1}{2} \left \langle \Bigl [S_{\bm{u^\prime},\bm{\tilde{u}}}^{(1),2}  \Bigr ]^2 \right \rangle$
& $-\int_{\bm{k},\Omega} k^4 \xi^2 F^{(l),(0)}_{\Omega,k}\Bigl ( \frac{(3\mu+\lambda)^2}{4}F^{(l),(0)}_{\Omega,k}$ &
$-\int_{\bm{k},\Omega}  k^4\xi^2 \Bigl ( \frac{(3\mu+2\lambda)^2}{2}[F^{(l),(0)}_{\Omega,k}]^2+\frac{(2\mu+\lambda)^2}{2}[F^{(t),(0)}_{\Omega,k}]^2$ \\
& $+ \frac{11\mu^2+10\mu\lambda+3\lambda^2}{8}F^{(t),(0)}_{\Omega,k}\Bigr )$ &$- \frac{(\mu+\lambda)^2}{4} F^{(l),(0)}_{\Omega,k}F^{(t),(0)}_{\Omega,k} \Bigr )$ \\
$- \frac{1}{2} \left \langle \Bigl [ S_{\bm{u^\prime},\tilde{h}}^{(1),2}\Bigr ]^2 \right \rangle$ &
$ -\frac{\mu^2\xi^2}{4}  \int_{\bm{k},\Omega} k^4 [G^{(0)}_{\Omega,k}]^2$&  $-\frac{(\mu+\lambda)^2\xi^2}{4}\int_{\bm{k},\Omega}  k^4 [G^{(0)}_{\Omega,k}]^2$
\end{tabular}
\end{ruledtabular}
\end{table}

The self-energies $\Sigma^{(t)}_{\omega,q}$ and $\Sigma^{(l)}_{\omega,q}$ determine renormalization of the
Lame coefficients:
\begin{equation}
\mu^\prime = \mu - \lim\limits_{q\to 0} \Bigl [ \Sigma^{(t)}_{\omega=0,q} - \Sigma_{\omega=0,q}\Bigr ]\Bigl /(\xi^2 q^2), \qquad
(\lambda+\mu)^\prime = \lambda+\mu - \lim\limits_{q\to 0} \Bigl [ \Sigma^{(l)}_{\omega=0,q}  -\Sigma_{\omega=0,q}\Bigr ]\Bigl /(\xi^2 q^2) .
\label{eq:defLame:i}
\end{equation}
We do not present the full expressions for the self-energies $\Sigma^{(l,t)}_{\omega,q}$ since they are too cumbersome. The linear in $k^2$ contributions to $\Sigma^{(l,t)}_{\omega=0,q}$ from different terms in Eq. \eqref{eq:dSu:i} are summarized in Table \ref{Tab1}. Using Eq. \eqref{eq:defLame:i} and Table \ref{Tab1}, we find
\begin{align}
\mu^\prime & = \mu -  \frac{1}{4} \int\limits_{\bm{k},\Omega} k^4
 \Biggl\{ d_{\rm c}  \mu^2 \bigl [ G^{(0)}_{\Omega,k}\bigr ]^2 + (3\mu+\lambda)^2 \bigl [ F^{(l),(0)}_{\Omega,k}]^2+  2\mu (\lambda+2\mu) F^{(l),(0)}_{\Omega,k}F^{(t),(0)}_{\Omega,k}\Biggr \} ,
 \label{eq:muren1-i} \\
(\lambda+\mu)^\prime & =\lambda+\mu -  \frac{1}{2} \int\limits_{\bm{k},\Omega} k^4
 \Biggl\{ d_{\rm c} (\mu+\lambda)^2 \bigl [ G^{(0)}_{\Omega,k}\bigr ]^2 + (3\mu+2\lambda)^2 \bigl [ F^{(l),(0)}_{\Omega,k}]^2 +  (2\mu+\lambda)^2  \bigl [ F^{(t),(0)}_{\Omega,k}]^2 \Biggr \} .
 \label{eq:lambdaren1-i}
\end{align}
\end{widetext}
Here we restore arbitrary value of $d_{\rm c}$.

Assuming that the infrared moment scale (which separates the slow and fast modes in the moment space) lies in the range $q_T < q < q_{\rm uv}\sim \sqrt{\min\{\lambda,\mu\}/\varkappa}$ we find that only the terms proportional to $d_{\rm c}$ provide logarithmically divergent contributions in
Eqs. \eqref{eq:muren1-i} and \eqref{eq:lambdaren1-i}. Hence, we find
\begin{equation}
\begin{split}
\frac{d\mu}{d\Lambda} & = - \frac{d_c}{32\pi \rho^{1/2}\varkappa^{3/2}} \mu^2 \ , \\
\frac{d\lambda}{d\Lambda} & = - \frac{d_c}{32\pi \rho^{1/2}\varkappa^{3/2}} (\mu^2+4\lambda\mu+2\lambda^2) \ .
\end{split}
\label{eq:Lame-ren-i}
\end{equation}
From Eqs. \eqref{eq:varkappa-ren-i} and \eqref{eq:Lame-ren-i} we obtain the renormalization group equations
\eqref{RG-Y} and \eqref{kappa0} of the main text. We see that $q_{\rm uv}\sim \sqrt{\min\{\lambda,\mu\}/\varkappa}$ is a natural ultraviolet cut-off  for the renormalization group equations \eqref{RG-Y} and \eqref{kappa0}.

As we mentioned in the main text,     the momentum  $ q_{\rm uv} \sim \sqrt{\min\{\lambda,\mu\}/\varkappa}$ is on the order of the inverse lattice constant $a^{-1}$ for graphene.  However, one can imagine a generic membrane, where $q_{\rm uv} \ll 1/a.$   Let us briefly discuss  what happens for  $ q_{\rm uv}< q < Q_{\rm uv} \sim1/a.$
Within this interval,
 there is no difference in the spectrum of in-plane and out-of-plane phonons. Then all terms in the right hand side of Eqs.  \eqref{eq:muren1-i} and \eqref{eq:lambdaren1-i} provide logarithmic contributions. Then we find the following renormalization group equations for the Lame coefficients in the range $q_{\rm uv} < q < Q_{\rm uv}$:
\begin{equation}
\begin{split}
\frac{d\mu}{d\tilde\Lambda} & = - \frac{(\lambda+4\mu)^2+(d_{\rm_c}-3)\mu^2}{32\pi \rho^{1/2}\varkappa^{3/2}} \ , \\
\frac{d\lambda}{d\tilde\Lambda} & = - \frac{2d_{\rm c}(\lambda+\mu)^2+(3\lambda+2\mu)^2 +(7-d_{\rm c})\mu^2}{32\pi \rho^{1/2}\varkappa^{3/2}} \ .
\end{split}
\label{eq:Lame-ren-2-i}
\end{equation}
where $\tilde\Lambda = \ln Q_{\rm uv}/q$. We note that in the range $q_{\rm uv} < q < Q_{\rm uv}$ there is no renormalization of the bending rigidity:
\begin{equation}
\frac{d\varkappa}{d\tilde\Lambda} =0  .
\end{equation}

%
%
%
%
%
%
%
%
%
%
%


\begin{thebibliography}{99}
\bibitem{comment}  Analyzing integral Eq.~\eqref{scl}, one can find that its low $\sigma$ asymplotics has form of the upper line in Eq.~\eqref{scl1} provided that
    $$C^{2(1-\eta)}=\left [\frac{A(\eta)}{2-\eta} \right] ^{2-\eta}\left (\frac{8B}{3Y} \right) ^{\eta} , $$
    where
    $$
    A(\eta)= \int_0^\infty \frac{\eta dy}{y^{1-\eta}(1+y^{2-\eta})}=\frac{2-\eta}{2}\Gamma\left(\frac{4-\eta}{2-\eta}\right)
    \Gamma\left(\frac{2-2\eta}{2-\eta}\right).
    $$
 We notice, that   for invariant subspace of elastic  moduli, $Y=8B/3,$  the coefficient  $C$  is temperature-independent.
 For $d_{\rm c} \to \infty $ ($\eta \to 0$),  we get $C=1/2.$
\end{thebibliography}
\end{document}